\documentclass[11pt,letterpaper]{article}
\usepackage[letterpaper,margin=1in]{geometry}
\usepackage{times}
\usepackage[T1]{fontenc}
\usepackage{amssymb}
\usepackage{mathtools}
\usepackage{graphicx}
\usepackage{array,booktabs,longtable}
\usepackage[expansion=false]{microtype}
\usepackage[numbers]{natbib}
\usepackage[colorlinks=true,allcolors=blue]{hyperref}
\usepackage{thm-restate}
\usepackage{thmtools}
\declaretheorem[name=Theorem]{theorem}
\declaretheorem[name=Lemma,sibling=theorem]{lemma}
\declaretheorem[name=Proposition,sibling=theorem]{proposition}
\declaretheorem[name=Corollary,sibling=theorem]{corollary}
\declaretheorem[name=Definition,sibling=theorem]{definition}
\declaretheorem[name=Remark,sibling=theorem]{remark}
\declaretheorem[name=Example,sibling=theorem]{example}
\newenvironment{proof}[1][Proof]
  {\par\noindent{\bfseries\upshape #1\ }}{\hfill$\blacksquare$\par}
\newenvironment{keywords}{\par\medskip\noindent\textbf{Keywords:}\ }{\par}
\newcolumntype{L}[1]{>{\raggedright\arraybackslash}p{#1}}
\DeclareMathOperator{\Bern}{Bern}
\DeclareMathOperator{\KL}{KL}

\newcommand{\EE}{\mathbb E}

\title{Optimal Conversion from Zero-Concentrated\\ to Gaussian Differential Privacy}

\author{Yu-Wei Chen\\
       Department of Statistics\\
       Purdue University\\
       West Lafayette, IN 47907, USA\\
       \texttt{chen4357@purdue.edu}
       \and
       Aaradhya Pandey\\
       Department of Statistics\\
       Harvard University\\
       Cambridge, MA 02138, USA\\
       \texttt{aaradhyapandey@fas.harvard.edu}
       \and
       Jordan Awan\\
       Department of Statistics\\
       University of Pittsburgh\\
       Pittsburgh, PA 15260, USA\\
       \texttt{jaa557@pitt.edu}}
\date{}

\begin{document}
\maketitle
\begin{abstract}
Gaussian differential privacy (GDP) and zero-concentrated differential privacy (zCDP) are two competing privacy accounting frameworks, but they share similar minimax lower bounds in various estimation problems. Moreover, it is known that a mechanism satisfying $\mu$-GDP satisfies $(\mu^2/2)$-zCDP, whereas the converse fails to hold at the same calibration. In this paper, we establish the optimal reverse conversion, showing that every $\rho$-zCDP mechanism satisfies $\mu_{\mathrm{zCDP}}(\rho)$-GDP, where $\mu_{\mathrm{zCDP}}(\rho)=2\Phi^{-1}((1+e^{-t_\rho})^{-1})$ and $t_\rho$ uniquely solves the equation $\rho=t_\rho\tanh(t_\rho/2)$. 
We also obtain simplified bounds $\sqrt{\pi\rho}\leq\mu_{\mathrm{zCDP}}(\rho)\leq\sqrt{8\rho}$, with sharp constants, which imply equivalence of the two frameworks up to constants.  This explains why their minimax risks agree up to a constant rescaling of the privacy budget. 
We show that the squared conversion is strictly convex, which implies that 
converting zCDP to GDP before applying composition or group privacy yields a tighter GDP guarantee than converting afterward. We also derive the sharp conversion from bounded-range and logistic DP mechanisms (e.g., exponential mechanism) to GDP, where converting before composition but after group privacy gives the tighter guarantee.
\end{abstract}

\begin{keywords}
R\'enyi divergence,  bounded-range mechanism, exponential mechanism, privacy composition, group privacy.
\end{keywords}

\clearpage

\section{Introduction}\label{sec:introduction}
Differential privacy (DP) limits what a randomized computation reveals about any individual's contribution, which creates plausible deniability and thereby provides privacy protection \cite{dwork2006calibrating,dwork2014algorithmic}. To satisfy DP, it requires the output distributions on neighboring datasets to be similar, and this similarity can be formalized either through divergences or through hypothesis testing. Concentrated differential privacy and R\'enyi differential privacy (RDP) bound divergences between the output distributions \cite{bun2016concentrated,mironov2017renyi}. The $f$-DP framework instead quantifies their similarity through hypothesis tests that distinguish them, requiring that at type I error $\alpha$, every such test has type II error at least $f(\alpha)$, for a given function $f$ \cite{dong2022gaussian}. Within this framework, Gaussian differential privacy (GDP) is the primary single-parameter framework, with $f$ given by testing between unit-variance Gaussian distributions with mean difference $\mu$.

Zero-concentrated differential privacy (zCDP) and GDP are both one-parameter frameworks motivated by the Gaussian mechanism, and their parameters have similar interpretations. Moreover, $\mu$-GDP implies $(\mu^2/2)$-zCDP, and the Gaussian mechanism attains both bounds exactly \cite{dong2022gaussian}. While zCDP is more commonly used in the computer science literature \cite{bun2016concentrated,kamath2019privately,biswas2020coinpress,canonne2020discrete}, where it offers additive composition, covers bounded-range mechanisms such as the exponential mechanism through standard conversions, and underlies the noisy measurements of the 2020 Census \cite{conversionAbowd2022}, GDP is more common in statistics \cite{dong2022gaussian, bu2020deep,awan2022logconcave,awan2023canonical,awan2024optimizing,wang2025bootstrap,chen2026near,pandey2025infinitely,pandey2026differential}, where its hypothesis-testing guarantee is directly interpretable, and \cite{dong2022gaussian} claim it is the tightest guarantee for the Gaussian mechanism, based on an argument on informativeness. Despite the similarities, it has not been established which GDP guarantee, if any, a zCDP budget implies, or whether the two frameworks are equivalent up to constants. Without such a result, a mechanism analyzed under the zCDP framework is hard to report in the more interpretable GDP framework of privacy accounting, and risk bounds proved in one framework cannot be transferred to the other.


The known implication from GDP to zCDP gives one direction, but its converse at the same calibration fails \cite{dong2022gaussian}. RDP-to-testing conversions derive trade-off functions from R\'enyi constraints, most recently the optimal one implied by RDP bounds at all orders \cite{riess2026optimal}, but they do not identify the sharp GDP parameter implied by a zCDP budget.

\subsection{Our results}\label{subsec:main-results}

In this paper, we establish the sharp zCDP-to-GDP conversion and use it to compare the mechanism classes and their minimax risks. We also derive privacy accounting consequences and a sharp bounded-range to GDP calculation. Specifically, our results are as follows:

\paragraph{Optimal conversion.}
Let $\Phi(t)$ be the cumulative distribution function (CDF) of a standard Gaussian. Then, for every $\rho>0$, the tightest (smallest) GDP parameter guaranteed by $\rho$-zCDP is
\[
\mu_{\mathrm{zCDP}}(\rho)=2\Phi^{-1}(\sigma(t_\rho)),
\]
where $t_\rho$ uniquely solves the calibration equation $\rho=t_\rho\tanh(t_\rho/2)$ and $\sigma(t)=(1+e^{-t})^{-1}$ is the logistic function. Our proof is by contrapositive, and the tightness of the conversion is achieved by the binary randomized response mechanism.

\begin{figure}[t]
    \centering
    \includegraphics[scale=0.89]{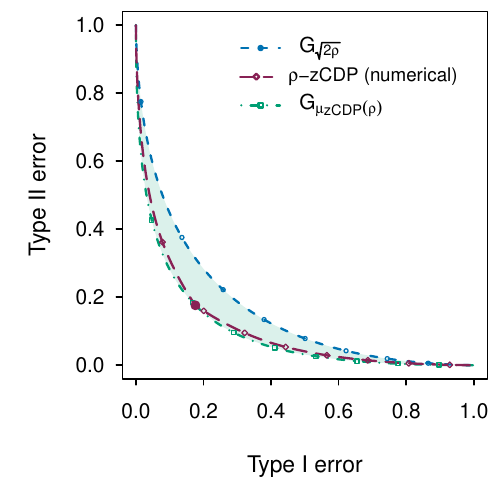}
    \caption{
    The three trade-off curves, $G_{\sqrt{2\rho}}$, the numerical $\rho$-zCDP testing boundary, and $G_{\mu_{\mathrm{zCDP}}(\rho)}$ illustrate the GDP--zCDP round-trip loss. 
    }
    \label{fig:zcdp-conversion}
\end{figure}

\paragraph{Conversion loss and statistical equivalence.}
We simplify the above conversion to obtain  $\sqrt{\pi\rho}\leq\mu_{\mathrm{zCDP}}(\rho)\leq\sqrt{8\rho}$, where $\mu_{\mathrm{zCDP}}(\rho)/\sqrt{\rho}$ increases from $\sqrt{\pi}$ to $\sqrt{8}$ over $\rho > 0$, so both constants are tight. Together with the known conversion from GDP to zCDP, this yields,
\begin{align*}
    \mu\text{-GDP}\ \implies\
({\mu^2}/{2})\text{-zCDP}\ \implies\ 2\mu\text{-GDP}.
\end{align*}

This shows that zCDP and GDP are equivalent up to constants. In particular, for any estimation problem, this yields a \emph{general minimax comparison} between GDP and zCDP up to constant-factor changes in the privacy budget.

\paragraph{Composition and group privacy.}
We prove that the squared conversion map $\rho\mapsto\mu_{\mathrm{zCDP}}(\rho)^2$ is strictly convex. This implies that converting individual zCDP budgets to GDP before applying composition or group privacy improves the resulting GDP bound.

\paragraph{GDP result for bounded-range and logistic mechanisms.}
Using similar techniques, we also prove the sharp conversion from logistic DP (including bounded-range mechanisms such as the exponential mechanism) to GDP. The square of the conversion map is strictly convex, favoring conversion to GDP before composition, while its concavity implies that group privacy should be applied before converting to GDP.

\begin{remark}
Figure~\ref{fig:zcdp-conversion} illustrates the conversion from $\rho$-zCDP to GDP at $\rho=1$. The trade-off function for $\rho$-zCDP is computed as follows:  For a type I error $\alpha$ and power $q$, let $K_{\mathrm{zCDP}}(\alpha,q)$ denote the smallest $\rho$ for which $\Bern(\alpha)$ and $\Bern(q)$, viewed as the output distributions of a mechanism on two neighboring datasets, satisfy $\rho$-zCDP. The figure shows the $\rho$-zCDP testing boundary $T_\rho^{\mathrm{zCDP}}(\alpha)=1-\sup\{q\in[\alpha,1]:K_{\mathrm{zCDP}}(\alpha,q)\le\rho\}$, which is the smallest type II error at level $\alpha$ among such Bernoulli pairs, computed numerically by solving $K_{\mathrm{zCDP}}(\alpha,q)=\rho$ for $q$ \cite{balle2020hypothesis,riess2026optimal}. This boundary lies above the curve $G_{\mu_{\mathrm{zCDP}}(\rho)}$, where $G_\mu$ gives the smallest type II error at each type I error level when testing $N(0,1)$ against $N(\mu,1)$, and touches it where the type I and type II errors both equal $\sigma(-t_\rho)$, so $\mu_{\mathrm{zCDP}}(\rho)$ cannot be decreased. The shaded region is the loss when a $\sqrt{2\rho}$-GDP guarantee, such as that of the Gaussian mechanism, is converted to $\rho$-zCDP and back.
\end{remark}

\subsection{Related work}

Divergence-based guarantees, such as zCDP and R\'enyi DP, control the moments of the privacy loss, which is the log-likelihood ratio between the output distributions on neighboring datasets \cite{bun2016concentrated,mironov2017renyi}. zCDP is equivalent to bounding the R\'enyi divergence at every order by a fixed multiple of the order, which is based on the R\'enyi divergence between two unit-variance Gaussians with differing means. Conversions from divergence-based guarantees to approximate DP, which bounds the likelihood ratio except on an event of small probability, have been established in \cite{bun2016concentrated,mironov2017renyi} and were sharpened by \cite{canonne2020discrete}. Pure DP, which bounds the privacy loss uniformly, converts sharply to RDP and zCDP, and binary randomized response attains these bounds \cite{steinke2024tight}. Exact zCDP budgets have been derived for several mechanisms, including bounded-range mechanisms, whose privacy loss lies in an interval of fixed width \cite{harrison2026exact,dong2020optimalcomposition}.

The $f$-DP framework, which includes GDP, was introduced by \cite{dong2022gaussian}, although previously the testing perspective has been used to study the composition of differential privacy mechanisms in \cite{kairouz2015composition}. 
\cite{dong2022gaussian} show that all popularly used divergence-based guarantees can be computed from $f$-DP without any loss. The reverse direction, from divergence bounds to testing errors, is studied by \cite{balle2020hypothesis,asoodeh2021variants}, where the latter also characterize the optimal approximate DP guarantee implied by an RDP bound, and by \cite{zhu2022optimal} through intersections of the sets of achievable type I and type II errors. The tightest $f$-DP guarantee implied by RDP bounds at all orders, including zCDP, has been characterized using Bernoulli experiments \cite{riess2026optimal}, while numerical methods have been developed for identifying and reporting the GDP guarantees of specific mechanisms \cite{liu2022bridge,gomez2026reporting}.

Appendix~\ref{app:privacy-conversions} provides additional context on the equations and formulae from the conversions discussed above. 


\section{Privacy definitions and randomized response}\label{sec:privacy-preliminaries}
Let $M$ be a Markov kernel (randomized mechanism) from an input measurable space to an output measurable space $(\Omega,\mathcal F)$, with a symmetric neighboring relation $\sim$ on its input space. For datasets of a fixed size, the usual choice is $x\sim x'$ when $d_{\mathrm H}(x,x')=1$, where $d_{\mathrm H}$ denotes the Hamming distance. For fixed neighboring datasets $x\sim x'$, write $P=M(x)$ and $Q=M(x')$. Let $\sigma(t)=(1+e^{-t})^{-1}$ denote the logistic function, and let $\Phi$ and $\varphi$ denote the standard normal CDF and density functions respectively.

\subsection{Privacy guarantees}\label{subsec:privacy-guarantees}

In this section, we define the two guarantees we compare, first zCDP and then GDP. For probability measures $P,Q$ dominated by a common measure $\nu$, with densities $p=dP/d\nu$ and $q=dQ/d\nu$, the R\'enyi divergence of order $\eta>1$ is $D_\eta(P\|Q)=\frac1{\eta-1}\log\int p^\eta q^{1-\eta}\,d\nu$ \cite{vanerven2014renyi}. Moreover, $\lim_{\eta\downarrow 1} D_\eta(P\| Q)$ results in the KL divergence. R\'enyi divergences are finite only if $Q(A)=0$ implies $P(A)=0$ for every measurable set $A$, denoted as the absolute continuity requirement $P\ll Q$. We say a mechanism $M$ satisfies $(\eta,\varepsilon)$-RDP if $D_\eta(M(x)\|M(x'))\le\varepsilon$ for every ordered neighboring pair \cite{mironov2017renyi}. Zero-concentrated DP (zCDP) requires such a bound at all $\eta>1$ simultaneously, with $\varepsilon=\rho\eta$ growing linearly in $\eta$.

\begin{definition}[Zero-concentrated DP \cite{bun2016concentrated}]\label{def:zCDP}
A mechanism $M$ satisfies $\rho$-zCDP if for every $x\sim x'$ and every $\eta>1$, $D_\eta(M(x)\|M(x'))\le\rho\eta$.
\end{definition}

This definition is motivated by the Gaussian mechanism. If $M(x)=f(x)+Z$ with $Z\sim N(0,\sigma_0^2 I)$, then $D_\eta(M(x)\|M(x')) =\eta\|f(x)-f(x')\|_2^2/(2\sigma_0^2)$. Thus, if $f$ has $\ell_2$-sensitivity $\Delta$, that is $\sup_{x\sim x'} \lVert f(x)-f(x')\rVert_2\leq \Delta$, then the mechanism satisfies $\rho$-zCDP with $\rho=\Delta^2/(2\sigma_0^2)$. 
Intuitively, a mechanism satisfies zCDP if it is ``more private'' than the Gaussian mechanism, in terms of the R\'enyi divergences. 
Smaller $\rho$ means that neighboring output distributions are closer, providing stronger privacy, and $\rho$-zCDP implies $\rho'$-zCDP for any $\rho'\geq \rho$.

For hypothesis testing, a measurable randomized test $\phi:\Omega\to[0,1]$ rejects $P$ with probability $\phi(\omega)$, based on the observation $\omega$. Its type I error and power are $\EE_P\phi$ and $\EE_Q\phi$. The trade-off function is $T(P,Q)(\alpha)=\inf_{\{\phi:\,\EE_P\phi\le\alpha\}}(1-\EE_Q\phi)$ for any $\alpha\in[0,1]$ \cite{torgersen1991comparison,dong2022gaussian}. The trade-off function returns the smallest achievable type II error at a given type I error level. Let $G_\mu(\alpha)=\Phi\bigl(\Phi^{-1}(1-\alpha)-\mu\bigr)$ be the trade-off function for $N(0,1)$ versus $N(\mu,1)$.

\begin{definition}[Gaussian DP \cite{dong2022gaussian}]\label{def:GDP}
A mechanism $M$ satisfies $\mu$-GDP if for every $x\sim x'$ and for every $\alpha\in[0,1]$, $T(M(x),M(x'))(\alpha)\ge G_\mu(\alpha)$.
\end{definition}

Equivalently, $\mu$-GDP holds when every test on neighboring outputs whose type I error and power lie in $(0,1)$ satisfies $\Phi^{-1}(\EE_Q\phi)-\Phi^{-1}(\EE_P\phi)\le\mu$. Replacing $G_\mu$ by a general trade-off function $f$ gives the $f$-DP framework. A smaller $\mu$ represents a stronger privacy guarantee with lower data usability, and $\mu$-GDP implies $\mu'$-GDP for any $\mu'\geq \mu$.

The Gaussian mechanism with variance $\sigma_0^2I$ applied to a function with $\ell_2$-sensitivity $\Delta$ satisfies $\Delta/\sigma_0$-GDP. Similar to zCDP, a mechanism is GDP if it is ``more private'' than the Gaussian mechanism, in terms of hypothesis testing.

\begin{proposition}[GDP implies zCDP \cite{dong2022gaussian}]\label{prop:gdp-zcdp}
For any $\mu>0$, if a mechanism satisfies $\mu$-GDP, then it also satisfies $(\mu^2/2)$-zCDP.
\end{proposition}

It is equivalent to state Proposition~\ref{prop:gdp-zcdp} as $\sqrt{2\rho}$-GDP implies $\rho$-zCDP for every $\rho>0$. However, the converse to Proposition \ref{prop:gdp-zcdp} at the same calibration fails. \cite[Proposition~B.7]{dong2022gaussian} give a binary randomized-response counterexample showing that $\rho$-zCDP does not generally imply $\sqrt{2\rho}$-GDP.

Both zCDP and GDP frameworks have concise rules for composition and group privacy. Suppose releases $M_1,\ldots,M_n$ are composed adaptively, so that each $M_i$ takes as input the dataset and the outputs of $M_1,\ldots,M_{i-1}$. For group privacy, consider groups of size $k$, that is, datasets $x^{(0)}$ and $x^{(k)}$ connected by a chain $x^{(0)}\sim x^{(1)}\sim\cdots\sim x^{(k)}$.
\begin{itemize}
    \item \emph{Composition.} If each $M_i$ is $\rho_i$-zCDP for every fixed realization of the preceding outputs, with deterministic $\rho_i$, then the joint release is $(\sum_{i=1}^n\rho_i)$-zCDP \cite{bun2016concentrated}. If each $M_i$ is instead $\mu_i$-GDP in the same sense, then the joint release is $\sqrt{\sum_{i=1}^n\mu_i^2}$-GDP \cite{dong2022gaussian}.
    \item \emph{Group privacy.} A $\rho$-zCDP mechanism satisfies $D_\eta(M(x^{(0)})\|M(x^{(k)}))\le k^2\rho\eta$ for every $\eta>1$ \cite{bun2016concentrated}, and a $\mu$-GDP mechanism satisfies $T(M(x^{(0)}),M(x^{(k)}))\ge G_{k\mu}$ \cite{dong2022gaussian}.
\end{itemize}

\subsection{Randomized Response}\label{subsec:rr}

For $t\ge0$, binary randomized response $\mathrm{RR}_t$ has input space and output space both as $\{0,1\}$. On input $x\in\{0,1\}$, it outputs $x$ directly with probability $\sigma(t)$ and outputs $1-x$ with probability $\sigma(-t)=1-\sigma(t)$.

Thus $\mathrm{RR}_t(0)\overset{d}{=}\Bern(\sigma(-t))$ and $\mathrm{RR}_t(1)\overset{d}{=}\Bern(\sigma(t))$, and $t=\log\{\sigma(t)/\sigma(-t)\}$ is the log odds of reporting the input correctly. At $t=0$ the output is a fair coin regardless of the input, and larger $t$ reports the input more accurately.

\begin{proposition}[Exact privacy parameters of randomized response]\label{prop:rr}
For any $t\ge0$, $\mathrm{RR}_t$ is tightly $t\tanh(t/2)$-zCDP and tightly $2\Phi^{-1}(\sigma(t))$-GDP. This tightness is in the sense that  no smaller parameter is valid in either framework.
\end{proposition}

The zCDP budget in Proposition~\ref{prop:rr} was derived in \cite{steinke2024tight,harrison2026exact}. Its value equals the KL limit $t\tanh(t/2)$ for this particular pair. The GDP budget follows from \cite[Lemma~2]{awan2024optimizing}. The GDP bound is tight at the test that takes the output as its guess of the input, which has type I and type II errors both equal to $\sigma(-t)$.

\section{Sharp conversion and its properties}\label{sec:sharp-conversions-accounting}

In this section, we present our main result (Theorem~\ref{thm:sharp-gdp-conversions}) on the sharp conversion from zCDP to GDP, its statistical consequences, and its shape and privacy accounting properties, followed by the GDP result for bounded-range and logistic mechanisms.

\subsection{The sharp zCDP-to-GDP conversion}\label{subsec:sharp-gdp-conversions}

The following theorem shows that every $\rho$-zCDP mechanism is $\mu_{\mathrm{zCDP}}(\rho)$-GDP. For every non-negative budget, this is the smallest universally valid GDP parameter.

\begin{restatable}[Sharp zCDP-to-GDP conversion]{theorem}{sharpconversion}\label{thm:sharp-gdp-conversions}
For $\rho\geq 0$, define
\begin{equation}\label{eq:murho_eqrho}
\mu_{\mathrm{zCDP}}(\rho)=2\Phi^{-1}(\sigma(t_\rho)),
\qquad
\rho=t_\rho\tanh(t_\rho/2),
\end{equation}
where $t_\rho\ge0$ uniquely solves the second equation.
Every $\rho$-zCDP mechanism is $\mu_{\mathrm{zCDP}}(\rho)$-GDP. Randomized response with parameter $t_\rho$ attains the bound by Proposition~\ref{prop:rr}.
\end{restatable}

\begin{figure}[t]
    \centering
    \includegraphics[scale=0.89]{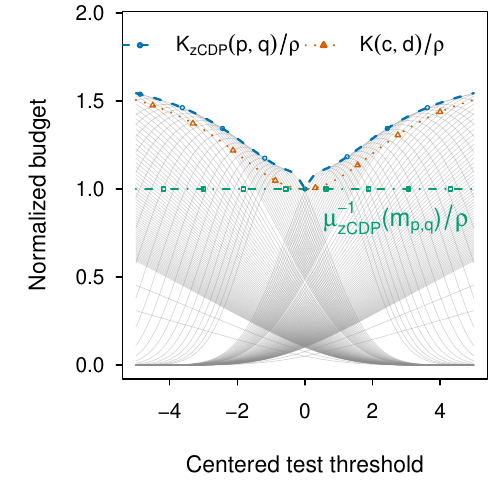}
    \caption{
    Comparison of zCDP budgets for Bernoulli pairs induced by Gaussian threshold tests. The horizontal axis is the centered threshold $\tau-\mu_{\mathrm{zCDP}}(\rho)/2$, and each budget on the vertical axis is divided by $\rho$. The exact budget $K_{\mathrm{zCDP}}(p,q)$ lies above the selected-order bound $K(c,d)$, which lies above the randomized-response budget $\rho$, shown as the horizontal line at $1$. Here $\rho=1$.
    }
    \label{fig:K}
\end{figure}

\begin{proof}[Proof sketch]
The proof is by contrapositive: Suppose a mechanism fails $\mu_{\mathrm{zCDP}}(\rho)$-GDP. Any optimal test results in a Bernoulli pair, and by data processing it suffices to show that this pair violates $\rho$-zCDP. Evaluating the R\'enyi divergence at a specific chosen order according to the pair, instead of the supremum over all orders, gives a closed-form lower bound on its zCDP budget and turns the claim into a one-dimensional (though non-trivial) calculus problem (Appendix~\ref{app:zcdp-conversion-proof}). Figure~\ref{fig:K} shows that this lower bound lies between the exact zCDP budget of the pair and the zCDP budget of randomized response with the same GDP parameter. Randomized response with parameter $t_\rho$ attains the bound by Proposition~\ref{prop:rr}.
\end{proof}

To further interpret Figure~\ref{fig:K}, consider the threshold test with cutoff value $\tau$ that rejects $N(0,1)$ in favor of $N(\mu_{\mathrm{zCDP}}(\rho),1)$ when the observation exceeds $\tau$. Its rejection probabilities are $p=1-\Phi(\tau)$ (type I error) and $q=1-\Phi(\tau-\mu_{\mathrm{zCDP}}(\rho))$ (power). At the midpoint threshold $\tau=\mu_{\mathrm{zCDP}}(\rho)/2$, the type~I and type~II errors are equal, so the induced Bernoulli pair coincides with symmetric randomized response. All three budgets therefore equal $\rho$ at this threshold. The figure centers the horizontal axis at this midpoint and divides each budget by $\rho$, so the three curves meet at $(0,1)$. A value above $1$ for either $K_{\mathrm{zCDP}}(p,q)/\rho$ or its lower bound $K(c,d)/\rho$ shows that the Bernoulli pair violates $\rho$-zCDP.

The next theorem describes the shape of the conversion $\mu_{\mathrm{zCDP}}$ from Theorem~\ref{thm:sharp-gdp-conversions}. Strict convexity of $\mu_{\mathrm{zCDP}}(\rho)^2$ implies that it is better to convert before applying composition or group privacy (Corollary~\ref{cor:composition-comparison}), and the bounds $\sqrt{\pi  \rho} \leq \mu_{\mathrm{zCDP}}(\rho)\leq \sqrt{8\rho}$ show that zCDP and GDP are equivalent up to constants.

\begin{restatable}[Shape and limiting scales of the zCDP to GDP conversion]{theorem}{conversionconvexity}\label{thm:conversion-shape}
The zCDP to GDP  conversion map  $\rho \mapsto \mu_{\mathrm{zCDP}}(\rho)$ is strictly increasing on $[0,\infty)$, and satisfies $\lim_{\rho\downarrow0}\mu_{\mathrm{zCDP}}(\rho)/\sqrt{\rho}=\sqrt{\pi}$ and $\lim_{\rho\to\infty}\mu_{\mathrm{zCDP}}(\rho)/\sqrt{\rho}=\sqrt{8}$. Moreover, $\rho \mapsto \mu_{\mathrm{zCDP}}(\rho)^2$ is strictly convex and $\rho \mapsto \mu_{\mathrm{zCDP}}(\rho)/\sqrt{\rho}$ is strictly increasing on $(0,\infty)$, so $\sqrt{\pi \rho} \leq \mu_{\mathrm{zCDP}}(\rho)\leq \sqrt{8\rho}$ for all $\rho\geq 0$.
\end{restatable}

\begin{proof}[Proof sketch]
Parameterizing the conversion by the randomized response parameter $t_\rho$ in Theorem~\ref{thm:sharp-gdp-conversions}, rather than by $\rho$, avoids differentiating the inverse Gaussian CDF and allows both limits to be taken in $t_\rho$, which tends to zero or infinity together with $\rho$. Showing that the inverse of $\mu_{\mathrm{zCDP}}^2$ is strictly increasing and strictly concave (Lemma~\ref{lem:gauss_integral_identity}) gives strict convexity of $\mu_{\mathrm{zCDP}}^2$, which with $\mu_{\mathrm{zCDP}}(0)=0$ gives the strict increase of $\mu_{\mathrm{zCDP}}(\rho)/\sqrt{\rho}$ (Appendix~\ref{app:conversion-properties-accounting}).
\end{proof}

Theorem~\ref{thm:conversion-shape} shows that zCDP and GDP are frameworks equivalent up to constants. Since $\mu_{\mathrm{zCDP}}(\rho)/\sqrt{\rho}$ increases from $\sqrt{\pi}$ to $\sqrt{8}$, every $\rho$-zCDP mechanism is $\sqrt{8\rho}$-GDP, and neither constant can be improved. Together with Proposition~\ref{prop:gdp-zcdp}, a $\mu$-GDP mechanism satisfies $(\mu^2/2)$-zCDP and converting back gives at worst $2\mu$-GDP, so passing through zCDP costs at most a factor of $2$ in the GDP parameter. Figure~\ref{fig:muzCDP2} in Appendix~\ref{app:conversion-properties-accounting} illustrates the shape and the limits of $\mu_{\mathrm{zCDP}}^2$ stated in Theorem~\ref{thm:conversion-shape}.

\begin{remark}[Minimax risks up to constants between zCDP and GDP]
Theorem~\ref{thm:conversion-shape} has a direct but significant consequence in statistics. Since every $\sqrt{2\rho}$-GDP estimator is $\rho$-zCDP and every $\rho$-zCDP estimator is $\mu_{\mathrm{zCDP}}(\rho)$-GDP, and hence $\sqrt{8\rho}$-GDP, the $\rho$-zCDP estimators lie between the $\sqrt{2\rho}$-GDP and the $\sqrt{8\rho}$-GDP estimators. For any estimation problem with a fixed sample size, the minimax risk under $\rho$-zCDP is therefore at most the minimax risk under $\sqrt{2\rho}$-GDP and at least the minimax risk under $\mu_{\mathrm{zCDP}}(\rho)$-GDP (Appendix~\ref{app:mechanism-classes-minimax}). Hence minimax lower bounds proved under GDP apply under zCDP, and risks achieved under GDP are achieved under zCDP, up to a constant rescaling of the privacy budget and regardless of the method used to prove them.
\end{remark}

Strict convexity of $\mu_{\mathrm{zCDP}}^2$ determines the order of accounting composition and group privacy in Corollary~\ref{cor:composition-comparison}, and Figure~\ref{fig:zcdp-accounting} in Appendix~\ref{app:conversion-properties-accounting} illustrates both comparisons in the corollary.

\begin{restatable}[Converting before accounting]{corollary}{convertbeforecomposing}\label{cor:composition-comparison}
For nonnegative $\rho_1,\ldots,\rho_n$ and $\rho$, and an integer $k\ge1$, 
\begin{enumerate}
    \item $\sqrt{\textstyle\sum_{i=1}^n\mu_{\mathrm{zCDP}}(\rho_i)^2}\le\mu_{\mathrm{zCDP}}\Bigl(\textstyle\sum_{i=1}^n\rho_i\Bigr)$ with strict inequality if at least two $\rho_i$ are positive.
    \item $k\mu_{\mathrm{zCDP}}(\rho)\le\mu_{\mathrm{zCDP}}(k^2\rho)$, with strict inequality when $\rho>0$ and $k>1$.
\end{enumerate}
\end{restatable}

\begin{proof}[Proof sketch]
Both inequalities follow from strict convexity of $\mu_{\mathrm{zCDP}}^2$ and $\mu_{\mathrm{zCDP}}(0)=0$ (Appendix~\ref{app:conversion-properties-accounting}).
\end{proof}

Both of the inequalities in Corollary~\ref{cor:composition-comparison} compare two ways of obtaining a GDP guarantee, and in each case the left side converts to GDP first. Corollary~\ref{cor:composition-comparison} provides direct consequences on the order of privacy accounting of composition and group privacy under zCDP and GDP. Under the adaptive composition introduced in Section~\ref{subsec:privacy-guarantees} with $\rho_i$-zCDP releases, composing in zCDP first gives $\mu_{\mathrm{zCDP}}(\sum_i\rho_i)$-GDP, while converting each release first and composing in GDP gives $\sqrt{\sum_i\mu_{\mathrm{zCDP}}(\rho_i)^2}$-GDP. Similarly, for a $\rho$-zCDP mechanism and groups of size $k$, applying zCDP group privacy first gives $\mu_{\mathrm{zCDP}}(k^2\rho)$-GDP, while converting first and applying GDP group privacy gives $k\mu_{\mathrm{zCDP}}(\rho)$-GDP. Corollary~\ref{cor:composition-comparison} therefore shows that converting first is superior in either case.

\subsection{Tight GDP for bounded-range and logistic mechanisms}\label{subsec:bounded-range-extension}
We next obtain a tight GDP guarantee for bounded-range and logistic mechanisms by converting directly to GDP rather than first through zCDP. First, we formally introduce these notions.

\begin{definition}[Bounded range]\label{def:bounded-range}
A mechanism satisfies $\varepsilon$-bounded range (BR) if, for every neighboring pair $P=M(x),Q=M(x')$, the distributions are mutually absolutely continuous and there is a real $\ell$ such that $\ell\le\log(dQ/dP)\le\ell+\varepsilon$ $P$-almost surely.
\end{definition}

\begin{figure}[t]
    \centering
    \includegraphics[scale=0.89]{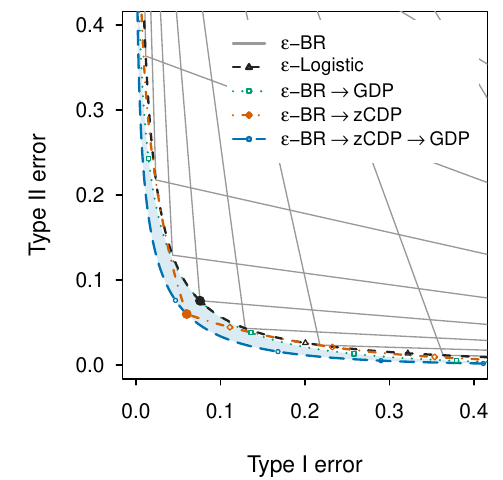}
    \caption{
    The five trade-off curves, BR examples, the logistic envelope, direct GDP conversion, the numerical zCDP testing boundary at $\rho_{\mathrm{BR}}(\varepsilon)$, and GDP conversion through zCDP, displayed on $[0,0.4]^2$. Here $\varepsilon=5$.
    }
    \label{fig:BR}
\end{figure}

Every $\varepsilon$-BR mechanism satisfies $(\varepsilon^2/8)$-zCDP. The sharper bound of \cite[Theorem~23]{harrison2026exact} is $\rho_{\mathrm{BR}}(\varepsilon)=\frac{\varepsilon}{e^\varepsilon-1}+\log\frac{e^\varepsilon-1}{\varepsilon}-1
$ for $\varepsilon>0,$ with $\rho_{\mathrm{BR}}(0)=0$. Although $\rho_{\mathrm{BR}}$ and $\mu_{\mathrm{zCDP}}$ are each sharp as conversions from BR to zCDP and from zCDP to GDP, respectively, it does not follow that their composition gives the sharp conversion from BR to GDP. Theorem~\ref{thm:br-gdp-conversion} establishes this direct, sharp conversion, while Figure \ref{fig:BR} illustrates that converting to GDP through zCDP is loose. 

\begin{example}[Bounded-range mechanisms]\label{ex:bounded-range}
(i) \emph{Exponential mechanism.} Let a real-valued score function $u(x,y)$ satisfy $|u(x,y)-u(x',y)|\le\Delta$ for all $x\sim x'$ and all outputs $y$. The mechanism with density (with respect to some common base measure) proportional to $\exp\{\varepsilon u(x,y)/(2\Delta)\}$ has $\log(dQ/dP)(y)=\frac{\varepsilon}{2\Delta}\{u(x',y)-u(x,y)\}+\log(Z_x/Z_{x'})$, where $Z_x$ is its normalizing constant. The first term lies in an interval of width $\varepsilon$ and the second does not depend on $y$, so the mechanism is $\varepsilon$-BR \cite{dong2020optimalcomposition}. (ii) \emph{Pure DP mechanisms} \cite{dwork2006calibrating}. A mechanism is $\varepsilon$-DP if $P(A)\le e^\varepsilon Q(A)$ for every neighboring pair $P=M(x),Q=M(x')$ and every measurable set $A$. Since the neighboring relation is symmetric, this is equivalent to $|\log(dQ/dP)|\le\varepsilon$ $P$-almost surely, so every $\varepsilon$-DP mechanism is $2\varepsilon$-BR. In particular, randomized response $\mathrm{RR}_\varepsilon$ from Section~\ref{subsec:rr} is $\varepsilon$-DP and hence $2\varepsilon$-BR.
\end{example}

Bounded-range mechanisms are members of a broader class of mechanisms, which satisfy logistic DP. Logistic DP is a convenient framework because it is closed under composition and group privacy.
\begin{definition}[Logistic differential privacy]\label{def:logistic-dp}
For $\varepsilon\ge0$, a mechanism is $\varepsilon$-logistic DP if every neighboring pair $P=M(x),Q=M(x')$ satisfies $T(P,Q)(\alpha)\ge L_\varepsilon(\alpha) :=\frac{1-\alpha}{1+(e^\varepsilon-1)\alpha}$ for all $0\le\alpha\le1$.
\end{definition}

Every $\varepsilon$-BR mechanism is $\varepsilon$-logistic DP \cite{dong2020gaussian}. For any randomized test with rejection probabilities $0<p<q<1$, bounded range implies $\log\{q(1-p)/(p(1-q))\}\le\varepsilon$, which is equivalent to $1-q\ge L_\varepsilon(p)$ (Appendix~\ref{app:bounded-range-conversion-proof}). The function $L_\varepsilon$ is the trade-off function between logistic distributions with locations $0$ and $\varepsilon$ and unit scale. This trade-off function arises from the additive logistic mechanism. If a statistic $T$ has sensitivity $\Delta$, then $T+(\Delta/\varepsilon)Z$ with $Z\sim\mathrm{Logistic}(0,1)$ satisfies $\varepsilon$-logistic DP, and the logistic noise is the unique log-concave canonical noise distribution for $\varepsilon$-logistic DP \cite{awan2022logconcave}. The trade-off function $L_\varepsilon(\alpha)$ is also the pointwise lower envelope of BR trade-off curves because, for each $0<p<1$, the $\varepsilon$-BR Bernoulli pair with $q=1-L_\varepsilon(p)$ attains $L_\varepsilon$ at $\alpha=p$. Thus BR remains our motivating class, while logistic DP is a broader class with convenient rules for adaptive composition and group privacy, which allows a common treatment of both. In particular, composition in logistic DP is additive in the $\varepsilon$ parameters, and a mechanism that satisfies $\varepsilon$-logistic DP satisfies $k\varepsilon$-logistic DP for groups of size $k$ \cite{dong2022gaussian,awan2022logconcave}. More details and formal statements about Logistic DP can be found in Appendix~\ref{app:logistic-background}.

\begin{restatable}[Sharp bounded-range-to-GDP conversion]{theorem}{brsharpconversion}\label{thm:br-gdp-conversion}
For $\varepsilon\ge0$, define
\begin{equation*}
\mu_{\mathrm{BR}}(\varepsilon)=2\Phi^{-1}(\sigma(\varepsilon/2)).
\end{equation*}
Every $\varepsilon$-BR mechanism is $\mu_{\mathrm{BR}}(\varepsilon)$-GDP. More generally, every $\varepsilon$-logistic DP mechanism is $\mu_{\mathrm{BR}}(\varepsilon)$-GDP. Tightness of both conversions is demonstrated by randomized response with parameter $\varepsilon/2$.
\end{restatable}

\begin{proof}[Proof sketch]
As in Theorem~\ref{thm:sharp-gdp-conversions}, we reduce to a Bernoulli pair. 
The BR parameter of a Bernoulli pair is the distance between the log-odds of its two success probabilities, and among Bernoulli pairs with BR budget $2t$, $\mathrm{RR}_t$ has the largest GDP parameter, so a pair failing $\mu_{\mathrm{BR}}(\varepsilon)$-GDP has BR budget above $\varepsilon$ (Appendix~\ref{app:bounded-range-conversion-proof}). The same argument covers logistic DP, whose constraint on a single test has the same log-odds bound. $\mathrm{RR}_{\varepsilon/2}$ attains the bound by Proposition~\ref{prop:rr}.
\end{proof}

For every $\varepsilon>0$, direct conversion $\mu_{\mathrm{BR}}(\varepsilon)$ is no looser than $\mu_{\mathrm{zCDP}}(\rho_{\mathrm{BR}}(\varepsilon))$ by the sharpness in Theorem~\ref{thm:br-gdp-conversion}. Figure \ref{fig:BR} also shows asymmetric Bernoulli BR examples, their pointwise lower envelope $L_\varepsilon$, and the numerical zCDP testing boundary at $\rho_{\mathrm{BR}}(\varepsilon)$, computed as above. The zCDP and GDP for BR are incomparable (numerically) because their curves cross each other. Both axes are restricted to $[0,0.4]$ for display; the guarantees hold over the full testing interval $[0,1]$. 

\begin{remark}[Recovery of the known pure-DP conversion]\label{rem:pure-dp-conversion}
By Example~\ref{ex:bounded-range}(ii), every $\varepsilon$-DP mechanism is $2\varepsilon$-BR. Theorem~\ref{thm:br-gdp-conversion} therefore recovers the sharp conversion of \cite[Lemma~2]{awan2024optimizing}, which states that every $\varepsilon$-DP mechanism is $\mu_{\mathrm{BR}}(2\varepsilon)$-GDP with $\mu_{\mathrm{BR}}(2\varepsilon)=2\Phi^{-1}(\sigma(\varepsilon))$, and $\mathrm{RR}_\varepsilon$ attains this parameter by Proposition~\ref{prop:rr}.
\end{remark}

\begin{restatable}[Shape and limiting scales of the bounded-range conversion]{theorem}{brconversionconvexity}\label{thm:br-conversion-shape}
The conversion map  $\varepsilon \mapsto \mu_{\mathrm{BR}}(\varepsilon)$ is strictly increasing and strictly concave on $[0,\infty)$, and satisfies $\lim_{\varepsilon\downarrow0}\mu_{\mathrm{BR}}(\varepsilon)/\varepsilon=\sqrt{\pi/8}$ and $\lim_{\varepsilon\to\infty}\mu_{\mathrm{BR}}(\varepsilon)/\sqrt{\varepsilon}=2$. Moreover, $\mu_{\mathrm{BR}}(\varepsilon)^2$ is strictly convex, $\mu_{\mathrm{BR}}(\varepsilon)^2/\varepsilon$ is strictly increasing, and $\mu_{\mathrm{BR}}(\varepsilon)/\varepsilon$ is strictly decreasing on $(0,\infty)$. It follows that $\mu_{\mathrm{BR}}(\varepsilon)\leq\min\{\sqrt{\pi/8}\,\varepsilon,\,2\sqrt{\varepsilon}\}$ for all $\varepsilon\geq0$, with both constants sharp by the two limits.
\end{restatable}

\begin{proof}[Proof sketch]
Since the bounded-range conversion $\mu_{\mathrm{BR}}(\varepsilon)$ is the GDP parameter $2\Phi^{-1}(\sigma(t))$ of randomized response $\mathrm{RR}_t$ at $t=\varepsilon/2$, as $\mu_{\mathrm{zCDP}}(\rho)$ is at $t=t_\rho$, the limits and the convexity of the square follow from the calculations for Theorem~\ref{thm:conversion-shape}, with $\varepsilon=2t$ in place of $\rho=t\tanh(t/2)$. Strict concavity follows because the derivative of $t\mapsto\Phi^{-1}(\sigma(t))$ is strictly decreasing on $[0,\infty)$ (Appendix~\ref{app:conversion-properties-accounting}).
\end{proof}

\begin{figure}[t]
    \centering
    \includegraphics[scale=0.89]{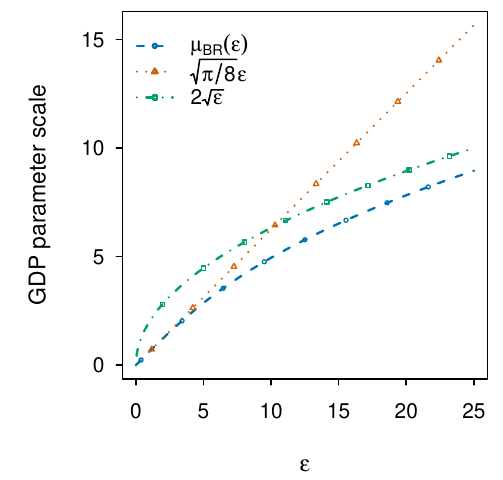}
    \caption{The sharp conversion $\varepsilon\mapsto\mu_{\mathrm{BR}}(\varepsilon)$ is strictly concave and is at most the minimum of $\sqrt{\pi/8}\,\varepsilon$ and $2\sqrt{\varepsilon}$.}
    \label{fig:muBR}
\end{figure}

The different limiting scales in Theorem~\ref{thm:conversion-shape} and Theorem~\ref{thm:br-conversion-shape} arise from how the privacy budgets parameterize the same randomized response mechanism $\mathrm{RR}_t$. Writing its GDP parameter as $2\Phi^{-1}(\sigma(t))$, the corresponding budgets are $\rho=t\tanh(t/2)$ and $\varepsilon=2t$. Near zero, the GDP parameter is linear in $t$, while $\rho$ is quadratic and $\varepsilon$ is linear, giving the respective scales $\sqrt{\rho}$ and $\varepsilon$. At infinity, the Gaussian quantile map composed with the sigmoid grows as $\sqrt{t}$, while both budgets grow linearly in $t$, yielding square-root scaling in both budgets. Figure~\ref{fig:muBR} illustrates the shape of $\mu_{\mathrm{BR}}$ and the limits of $\mu_{\mathrm{BR}}$ stated in Theorem~\ref{thm:br-conversion-shape}, and Figure~\ref{fig:muBR2} in Appendix~\ref{app:conversion-properties-accounting} illustrates the shape of $\mu_{\mathrm{BR}}^2$.

\begin{restatable}[Order of conversion and accounting for logistic DP]{corollary}{brconvertbeforecomposing}\label{cor:br-composition-comparison}
For nonnegative $\varepsilon_1,\ldots,\varepsilon_n$ and $\varepsilon$, and an integer $k\ge1$, 
\begin{enumerate}
    \item $\sqrt{\textstyle\sum_{i=1}^n\mu_{\mathrm{BR}}(\varepsilon_i)^2}\le\mu_{\mathrm{BR}}\Bigl(\textstyle\sum_{i=1}^n\varepsilon_i\Bigr)$ with strict inequality if at least two $\varepsilon_i>0$.
    \item $\mu_{\mathrm{BR}}(k\varepsilon)\le k\mu_{\mathrm{BR}}(\varepsilon)$ with strict inequality if $\varepsilon>0$ and $k>1$.
\end{enumerate}
\end{restatable}

\begin{proof}[Proof sketch]
The first inequality follows from strict convexity of $\mu_{\mathrm{BR}}^2$, and the second from strict concavity of $\mu_{\mathrm{BR}}$, along with $\mu_{\mathrm{BR}}(0)=0$ for each (Appendix~\ref{app:conversion-properties-accounting}).
\end{proof}

Corollary~\ref{cor:br-composition-comparison} provides direct consequences on the order of privacy accounting of composition and group privacy under logistic DP and GDP. Under the adaptive composition of $\varepsilon_i$-logistic DP releases, composing in logistic DP first gives $\mu_{\mathrm{BR}}(\sum_i\varepsilon_i)$-GDP, while converting each release first and composing in GDP gives $\sqrt{\sum_i\mu_{\mathrm{BR}}(\varepsilon_i)^2}$-GDP, so converting first gives the tighter result. For groups of size $k$, applying logistic group privacy first gives $\mu_{\mathrm{BR}}(k\varepsilon)$-GDP, while converting first and applying GDP group privacy gives $k\mu_{\mathrm{BR}}(\varepsilon)$-GDP, so applying group privacy first is tighter. Appendix~\ref{app:logistic-background} introduces the accounting properties and their connection to the log-concavity of the logistic density. 

Since every $\varepsilon$-BR mechanism is $\varepsilon$-logistic DP, both inequalities apply to BR mechanisms. Figure~\ref{fig:br-accounting} in Appendix~\ref{app:conversion-properties-accounting} illustrates Corollary~\ref{cor:br-composition-comparison} by plotting the ratio of the quantities in each inequality.

\section{Discussion}\label{sec:discussion}
We established the sharp conversions from zCDP and from bounded-range and logistic DP to GDP. We also established that zCDP and GDP are equivalent up to constants. Our conversions allow for more mechanisms to be analyzed under GDP, which is growing in popularity \cite{gomez2026reporting}. Since it is superior to convert from zCDP to GDP before applying either composition or group privacy, this offers additional evidence that GDP is a natural framework for DP. 

Our conversions hold for any generic mechanism with only a given zCDP or BR budget, so direct GDP analysis of specific mechanisms may be tighter. Likewise, when R\'enyi bounds at all orders are known, the optimal trade-off guarantee of \cite{riess2026optimal} can be tighter than the Gaussian curve implied by the zCDP budget alone. Our analysis covers only zCDP, bounded-range DP, and logistic DP, and other mechanisms or other divergence-based guarantees are outside the scope of this paper.

Several questions remain open. The exception for group privacy under bounded-range DP has a simple algebraic reason, since group privacy scales the zCDP budget by the square of the group size and the bounded-range budget by the group size itself, but an intuitive explanation is unclear. Better understanding could clarify a general rule for when early conversion is or is not beneficial for a general privacy guarantee. The discrete Gaussian mechanism has a known zCDP guarantee \cite{canonne2020discrete}, which our result converts to GDP, but to our knowledge no direct GDP guarantee is currently available for this mechanism, and such an analysis may improve the converted parameter and offer insight into the privacy guarantees of the 2020 U.S. Decennial Census products. Finally, a bound at a single finite R\'enyi order does not generally imply any finite GDP guarantee; identifying useful additional conditions that permit such a conversion is a direction for future work.

\section*{Acknowledgment}
This work was supported in part by NSF award SES-2610910 to University of Pittsburgh. 

\textbf{AI Disclosure.} The authors derived the series expansion in the proof of Lemma~\ref{lem:quantile-ratio}, and used Wolfram|Alpha and ChatGPT to double-check its correctness. Furthermore, the authors used ChatGPT to polish the wording of the manuscript and develop function visualizations from author-written code.

\newpage

\appendix
\section{Additional related work and mechanism trade-off curves}\label{app:privacy-conversions}

\subsection{Related work}

The standard conversion of \cite{mironov2017renyi} states that $(\eta,\widehat\varepsilon)$-RDP implies
\begin{equation}\label{eq:known-rdp-adp}
\left(\widehat\varepsilon+\frac{\log(1/\delta)}{\eta-1},\delta\right)\text{-DP},
\end{equation}
for $\delta\in(0,1)$. Applying \eqref{eq:known-rdp-adp} with $\widehat\varepsilon=\rho\eta$ at each order $\eta>1$ and optimizing over $\eta$ gives the standard implication of \cite{bun2016concentrated}, that is, if a mechanism is $\rho$-zCDP, then it satisfies
\begin{equation}\label{eq:known-zcdp-adp}
\left(\rho+2\sqrt{\rho\log(1/\delta)},\delta\right)\text{-DP}.
\end{equation}

The conversions in \eqref{eq:known-rdp-adp}--\eqref{eq:known-zcdp-adp} are not generally optimal. Optimal approximate-DP guarantees from a given RDP constraint have been characterized in \cite{asoodeh2021variants}, while an improved conversion that can also be optimized over R\'enyi orders is given in \cite{canonne2020discrete}. $(\varepsilon,\delta)$-DP is shown to be equivalent to $f_{\varepsilon,\delta}$-DP \cite{dong2022gaussian}, where $f_{\varepsilon,\delta}(\alpha)=\max\{0,1-\delta-e^\varepsilon\alpha, e^{-\varepsilon}(1-\delta-\alpha)\}$. GDP corresponds exactly to the family $(\varepsilon,\delta_\mu(\varepsilon))$-DP for all $\varepsilon\geq0$ \cite{balle2018improving}, where
\begin{equation*}
\delta_\mu(\varepsilon)
=\Phi\!\left(\frac\mu2-\frac\varepsilon\mu\right)
-e^\varepsilon\Phi\!\left(-\frac\mu2-\frac\varepsilon\mu\right).
\end{equation*}
The divergence comparison in \cite{dong2022gaussian} gives the GDP-to-zCDP implication stated in Proposition~\ref{prop:gdp-zcdp}.

\subsection{Illustration of mechanism trade-off curves}

\begin{figure}[t]
    \centering
    \includegraphics[scale=0.89]{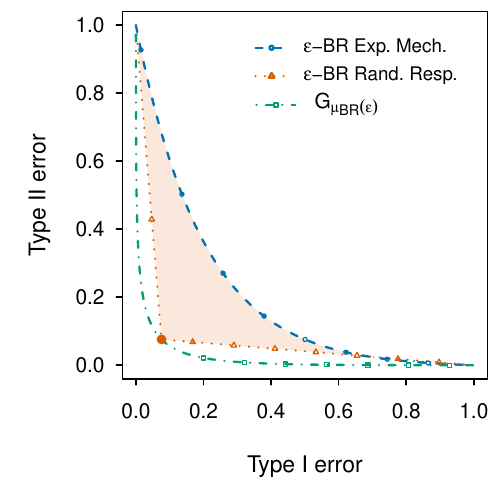}
    \includegraphics[scale=0.89]{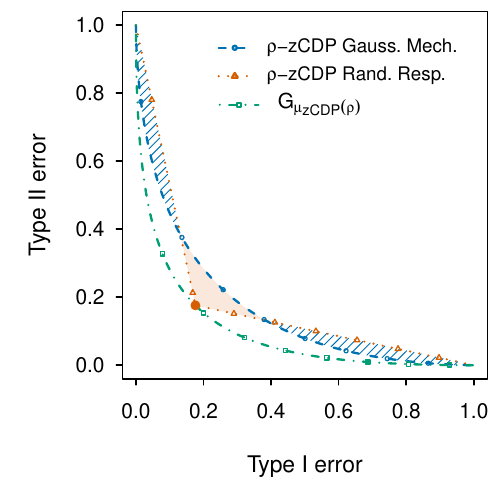}
    \caption{Left: $T_{\mathrm{EM}}(\alpha)=(e^{\varepsilon(1-\alpha)}-1)/(e^\varepsilon-1)$ and $T_{\mathrm{RR}}(\alpha)=\max\{1-e^{\varepsilon/2}\alpha,e^{-\varepsilon/2}(1-\alpha)\}$ lie above $G_{\mu_{\mathrm{BR}}(\varepsilon)}$ with $\varepsilon=5$. Right: trade-off curves of the Gaussian mechanism and binary randomized response satisfying $\rho$-zCDP with $\rho=1$. 
    }
    \label{fig:zcdp-BR-app}
\end{figure}

The left panel of Figure~\ref{fig:zcdp-BR-app} compares densities proportional to $e^{\varepsilon xy}$ on $[0,1]$ with Bernoulli success probabilities $\sigma(\varepsilon x-\varepsilon/2)$ for neighboring inputs $x\in\{0,1\}$. For the exponential mechanism, $p(y\mid x=0)=1$ and $p(y\mid x=1)=e^{\varepsilon y}\varepsilon/(e^\varepsilon-1)$. Thus, the type II error $T_{\mathrm{EM}}(\alpha)=\int_0^{1-\alpha}p(y\mid x=1)\,dy=(e^{\varepsilon(1-\alpha)}-1)/(e^\varepsilon-1)$ and the log density ratio $\log \left[p(y\mid x=1)/p(y\mid x=0)\right] =\varepsilon y+\log \left[\varepsilon/(e^\varepsilon-1)\right]$. For randomized response, $p(y=1\mid x=0)=p(y=0\mid x=1)=1/(1+e^{\varepsilon/2})$ and $p(y=1\mid x=1)=p(y=0\mid x=0)=e^{\varepsilon/2}/(1+e^{\varepsilon/2})$. The most powerful test first rejects at $Y=1$, randomizing to attain size $\alpha$. For $\alpha\le p(y=1\mid x=0)$, this gives type II error $1-\alpha p(y=1\mid x=1)/p(y=1\mid x=0)=1-e^{\varepsilon/2}\alpha$; for $\alpha>p(y=1\mid x=0)$, it also rejects at $Y=0$, giving $p(y=0\mid x=1)(1-\alpha)/p(y=0\mid x=0)=e^{-\varepsilon/2}(1-\alpha)$. Hence $T_{\mathrm{RR}}(\alpha)=\max\{1-e^{\varepsilon/2}\alpha,e^{-\varepsilon/2}(1-\alpha)\}$. Consequently, both mechanisms have exact $\varepsilon$-BR, but their trade-off curves cross, so neither uniformly dominates. Orange shading and blue hatching indicate where the trade-off curve of the binary randomized response $T_{\mathrm{RR}}$ and that of the exponential mechanism $T_{\mathrm{EM}}$, respectively, are lower. Randomized response touches the Gaussian lower bound at its fixed (equal-error) point, establishing sharpness of the GDP parameter.

Similarly, the right panel compares the trade-off curves between $\rho$-zCDP Gaussian mechanism and randomized response mechanism. Orange shading and blue hatching indicate where the trade-off curve of the binary randomized response and that of the Gaussian mechanism, respectively, are lower. Again, randomized response touches the Gaussian lower bound at its fixed (equal-error) point, establishing sharpness of the GDP parameter.

\section{Proof of the sharp zCDP-to-GDP conversion}\label{app:zcdp-conversion-proof}

In the zCDP case, a failed GDP guarantee means
\begin{equation}\label{eq:gdp-failure}
T(P,Q)(\alpha)<G_{\mu_{\mathrm{zCDP}}(\rho)}(\alpha)
\end{equation}
for some neighboring pair of output distributions $P,\,Q$ and at some testing level $\alpha$. Thus starting with \eqref{eq:gdp-failure}, the desired contradiction should end with
\begin{equation}\label{eq:desired-violation}
D_\eta(P\|Q)>\rho\eta\quad\text{or}\quad D_\eta(Q\|P)>\rho\eta
\end{equation}
for some order $\eta>1$.

\subsection{Bernoulli reduction}\label{app:bernoulli-reduction}

Let $\mu>0$, and suppose that $T(P,Q)(\alpha)<G_\mu(\alpha)$ for some $P=M(x),\,Q=M(x')$ with neighboring $x \sim x'$. Then there exists a measurable test $\phi$ such that $p:=\EE_P\phi\leq\alpha$ and $1-q:=1-\EE_Q\phi<G_\mu(\alpha)$.

The identity $\Phi^{-1}(1-\alpha)=-\Phi^{-1}(\alpha)$ gives $1-G_\mu(\alpha)
=1-\Phi\bigl(-\Phi^{-1}(\alpha)-\mu\bigr)
=\Phi\bigl(\Phi^{-1}(\alpha)+\mu\bigr)$. Since $p\leq\alpha$, monotonicity implies
\begin{equation}\label{eq:test-power-violation}
\EE_Q\phi = q>\Phi\bigl(\Phi^{-1}(\alpha)+\mu\bigr)
\geq\Phi\bigl(\Phi^{-1}(p)+\mu\bigr).
\end{equation}

Thus violation of $\mu$-GDP yields a test $\phi$ satisfying \eqref{eq:test-power-violation}.
Here $\Phi^{-1}(0)=-\infty$ and $\Phi^{-1}(1)=+\infty$, with $\Phi(-\infty)=0$ and $\Phi(+\infty)=1$, so \eqref{eq:test-power-violation} also covers $p=0$. The cases $p=0$ and $q=1$ are treated after \eqref{eq:data-processing}. For interior probabilities $0<p<q<1$, solving $q=\Phi\bigl(\Phi^{-1}(p)+m\bigr)$ for $m$ gives $m_{p,q}:=\Phi^{-1}(q)-\Phi^{-1}(p)$. The quantity $m_{p,q}$ is the Gaussian mean separation that reproduces type I error $p$ and power $q$. Equation~\eqref{eq:test-power-violation} implies a key inequality
\begin{align*}
m_{p,q}=\Phi^{-1}(q)-\Phi^{-1}(p)>\mu.
\end{align*}

\subsubsection{The binary decision and data processing}

The benefit of using testing formulation is as follows. Given any mechanism output $\omega \in \Omega$, one can come up with a binary decision $B\mid\omega\sim\Bern(\phi(\omega))$, where $B=1$ favors $x'$ and $B=0$ favors $x$. Averaging over the output distribution gives
\begin{align*}
\Pr(B=1\mid\text{input}=x)
&=\int_\Omega\phi(\omega)\,P(d\omega)=\EE_P\phi=p,\\
\Pr(B=1\mid\text{input}=x')
&=\int_\Omega\phi(\omega)\,Q(d\omega)=\EE_Q\phi=q.
\end{align*}

Thus $B$ has distribution $P'=\Bern(p)$ under input $x$ and distribution $Q'=\Bern(q)$ under input $x'$. Since this construction applies the same randomized post-processing to both output distributions, the R\'enyi data-processing inequality \cite[Theorem~9 and Example~2]{vanerven2014renyi} gives, for every $\eta>1$,
\begin{equation}\label{eq:data-processing}
D_\eta(P'\|Q')\leq D_\eta(P\|Q),\qquad D_\eta(Q'\|P')\leq D_\eta(Q\|P).
\end{equation}

Here $P=M(x)$ and $Q=M(x')$. Consequently, if either of \eqref{eq:data-processing} exceeds $\rho\eta$, that corresponding pair would have divergence exceeding $\rho\eta$. For boundary cases, \eqref{eq:test-power-violation} implies $p<q$. If $p=0<q$, then $D_\eta(Q'\|P')=\infty$ for every $\eta>1$. If $p<q=1$, then $D_\eta(P'\|Q')=\infty$. By \eqref{eq:data-processing}, these cases immediately give \eqref{eq:desired-violation}. It therefore remains to consider $0<p<q<1$.

\subsubsection{Bernoulli divergence}

For a Bernoulli pair,
\begin{align}\label{eq:bernoulli-renyi}
D_\eta(\Bern(p)\|\Bern(q))
=\frac{1}{\eta-1}\log\left[q\left(\frac{p}{q}\right)^\eta+(1-q)\left(\frac{1-p}{1-q}\right)^\eta\right].
\end{align}

Note that for $\eta>1$, the bound $D_\eta\le\rho\eta$ is equivalently a bound of $e^{\rho\eta(\eta-1)}$ on the expression inside the logarithm. Define
\begin{equation}\label{eq:H-definition}
K_{\mathrm{zCDP}}(p,q):=\sup_{\eta>1}\frac{1}{\eta}\max\{D_\eta(\Bern(p)\|\Bern(q)),\,D_\eta(\Bern(q)\|\Bern(p))\}.
\end{equation}

By \eqref{eq:data-processing}, it suffices to show that the binary pair obtained from a GDP violation satisfies $K_{\mathrm{zCDP}}(p,q)>\rho$. 

\subsubsection{From Gaussian separation to zCDP budget}

For $0<p<q<1$, write $m_{p,q}:=\Phi^{-1}(q)-\Phi^{-1}(p)$. The implication we need becomes
$$
m_{p,q}>\mu_{\mathrm{zCDP}}(\rho) \implies K_{\mathrm{zCDP}}(p,q)>\rho.
$$ 

To compare a separation with a budget, we invert $\mu_{\mathrm{zCDP}}$. For a separation $m\ge0$, let 
\begin{equation}\label{eq:inverse-common-parameter}
a=\log\{\Phi(m/2)/\Phi(-m/2)\}, 
\end{equation}
so that $2\Phi^{-1}(\sigma(a))=m$. Setting $t_\rho=a$ in \eqref{eq:murho_eqrho} gives $\mu_{\mathrm{zCDP}}(a\tanh(a/2))=2\Phi^{-1}(\sigma(a))=m$. Since $\sigma(a)=\Phi(m/2)$ and $\tanh(a/2)=2\sigma(a)-1$, the inverse conversion for $m\ge 0$ is
\begin{equation}\label{eq:R-definition}
\mu_{\mathrm{zCDP}}^{-1}(m):=(2\Phi(m/2)-1)\log\frac{\Phi(m/2)}{\Phi(-m/2)}
\end{equation}

Both factors in \eqref{eq:R-definition} equal zero at $m=0$ and have positive derivatives $\varphi(m/2)$ and $\frac{\varphi(m/2)}2[1/\Phi(m/2)+1/\Phi(-m/2)]$, so $\mu_{\mathrm{zCDP}}^{-1}$ is strictly increasing on $[0,\infty)$. Once Lemma~\ref{lem:bernoulli-bound} is established, applying \eqref{eq:R-definition} at $m=m_{p,q}$ shows that a GDP violation gives
\[
K_{\mathrm{zCDP}}(p,q)\ge\mu_{\mathrm{zCDP}}^{-1}(m_{p,q})
>\mu_{\mathrm{zCDP}}^{-1}(\mu_{\mathrm{zCDP}}(\rho))=\rho,
\]
contradicting zCDP by data processing. 

\subsection{The selected R\'enyi order}\label{app:log-odds-selected-order}

A direct approach to \eqref{eq:bernoulli-target} would write $p=\Phi(x)$ and $q=\Phi(x+m)$, substituting into \eqref{eq:bernoulli-renyi} to get $D_\eta(\Bern(p)\|\Bern(q))=\frac{1}{\eta-1}\log[\Phi(x)^\eta\Phi(x+m_{p,q})^{1-\eta}+\Phi(-x)^\eta\Phi(-x-m_{p,q})^{1-\eta}]$. This is very cumbersome to deal with. 

Instead, we use log-odds coordinates for \eqref{eq:bernoulli-renyi} as $u:=\log\frac{p}{1-p}$ and $v:=\log\frac{q}{1-q}$ with $u<v$, so $p=\sigma(u)$ and $q=\sigma(v)$. For the convenience of communication, we define
\begin{equation*}
g(t)=\Phi^{-1}(\sigma(t)).
\end{equation*}
The Gaussian separation remains $m_{p,q}=g(v)-g(u)$.

Since $p=e^u/(1+e^u)$ and $1-p=1/(1+e^u)$, and similarly for $q$ with $v$, the terms $p^\eta q^{1-\eta}$ and $(1-p)^\eta(1-q)^{1-\eta}$ in \eqref{eq:bernoulli-renyi} equal $e^{\eta u+(1-\eta)v}$ and $1$, each divided by $(1+e^u)^\eta(1+e^v)^{1-\eta}$. Their sum is therefore $(1+e^{\eta u+(1-\eta)v})/[(1+e^u)^\eta(1+e^v)^{1-\eta}]$. With $A(s):=\log(1+e^s)$ and $h(s):=\log\cosh(s/2)$, so that $A(s)=s/2+\log 2+h(s)$, taking logarithms gives
\begin{equation}\label{eq:exponential-family-renyi}
\begin{aligned}
(\eta-1)D_\eta(\Bern(p)\|\Bern(q))
&=A(\eta u+(1-\eta)v)-\eta A(u)-(1-\eta)A(v)\\
&=h(\eta u+(1-\eta)v)-\eta h(u)-(1-\eta)h(v).
\end{aligned}
\end{equation}
The second equality holds because the affine part $s/2+\log 2$ contributes zero. 

Since $h$ is even, a suitable choice of $\eta$ simplifies this expression. First suppose $u+v>0$. In \eqref{eq:exponential-family-renyi}, $\eta$ enters $h$ only through the argument $v-\eta(v-u)$ of the first term. We choose $\eta$ so that this equals $-v$, allowing us to use $h(-v)=h(v)$. Solving $v-\eta(v-u)=-v$ gives
\begin{equation*}
\eta_*=\frac{2v}{v-u} = 1+\frac{v+u}{v-u},
\end{equation*}
and $\eta_*>1$ because $u+v>0$ and $v-u>0$. At this specific order, $(\eta_*-1)D_{\eta_*}(\Bern(p)\|\Bern(q))=h(-v)-\eta_* h(u)-(1-\eta_*)h(v)=\eta_*[h(v)-h(u)]$, which reduces to 
$$
D_{\eta_*}(\Bern(p)\|\Bern(q))/\eta_*=(v-u)(h(v)-h(u))/(u+v).
$$ 
If $u+v<0$, replacing $(p,q)$ by $(1-q,1-p)$ maps $(u,v)$ to $(-v,-u)$. This preserves $K_{\mathrm{zCDP}}$ and the Gaussian separation and makes $u+v>0$. We may therefore assume $u+v\ge 0$. The case $u+v=0$ corresponds to the continuous extension of $K$ at $c=0$ described after \eqref{eq:K-lower-bound}.

For convenience, we introduce $c:=(u+v)/2\ge 0$ and $d:=(v-u)/2>0$, so $u=c-d$, $v=c+d$, $p=\sigma(c-d)$, and $q=\sigma(c+d)$. Symmetry corresponds to $c=0$. For $c>0$, the selected order and its lower bound become $\eta_*=1+c/d$ and
\begin{equation}\label{eq:K-lower-bound}
K_{\mathrm{zCDP}}(p,q)\ge\frac{D_{\eta_*}(\Bern(p)\|\Bern(q))}{\eta_*}=K(c,d):=\frac{d}{c}[h(c+d)-h(c-d)].
\end{equation}
The corresponding Gaussian separation is $m_{p,q}=g(c+d)-g(c-d)$. Note that $g$ is odd (since $\sigma(-s)=1-\sigma(s)$ and $\Phi$ is symmetric), with $g'(s)=\sigma'(s)/\varphi(g(s))>0$.

At $c=0$, we have $p=\sigma(-d)$, $q=\sigma(d)=1-p$, and $\eta_*=1$, where the formula for $K$ is undefined. The R\'enyi divergence at order one is the KL divergence, $\KL(\Bern(\sigma(-d))\|\Bern(\sigma(d)))=p(-d)+(1-p)d=d\tanh(d/2)$. This agrees with the limit of $K$. By evenness of $h$, $h(c+d)-h(c-d)=c\int_{-1}^{1}h'(d+ct)\,dt$, so $K(c,d)=d\int_{-1}^{1}h'(d+ct)\,dt$, which tends to $2dh'(d)=d\tanh(d/2)$ as $c\to0$. We therefore set $K(0,d)=d\tanh(d/2)$.

\subsection{Symmetrization at fixed Gaussian separation}\label{app:symmetrization}

Equation~\eqref{eq:K-lower-bound} gives $K_{\mathrm{zCDP}}(p,q)\ge K(c,d)$, so \eqref{eq:bernoulli-target} follows from
\begin{equation}\label{eq:K-target}
K(c,d)\ge \mu_{\mathrm{zCDP}}^{-1}(g(c+d)-g(c-d)).
\end{equation}
Let $a>0$ solve $2g(a)=g(c+d)-g(c-d)$, which has a unique solution because $g$ is strictly increasing from zero to infinity on $[0,\infty)$. The symmetric pair with log-odds $-a,a$ has the same Gaussian separation, and the derivation of \eqref{eq:R-definition} with $m=2g(a)$ gives
\begin{equation}\label{eq:R-symmetric}
\mu_{\mathrm{zCDP}}^{-1}(2g(a))=a\tanh(a/2)=K(0,a).
\end{equation}
Thus \eqref{eq:K-target} is equivalent to $K(c,d)\ge K(0,a)$, that is, among pairs with the same Gaussian separation, the symmetric pair minimizes $K$.

To prove this, we fix the separation $2g(a)$ and vary the midpoint $c$, choosing the half-distance $d(c)$ so that $g(c+d(c))-g(c-d(c))=2g(a)$ with $d(0)=a$. As $c$ grows, the interval $[c-d(c),c+d(c)]$ moves toward the tail where $g'$ is smaller, so $d(c)$ must grow to keep the separation, and the two comparisons below show that this growth keeps $K$ from decreasing. Then, it suffices to show that $c\mapsto K(c,d(c))$ is non-decreasing. The proof of Lemma~\ref{lem:symmetrization} in Appendix~\ref{app:symmetrization} shows that, with $x=c-d(c)$ and $y=c+d(c)$, this holds if
\begin{equation}\label{eq:two-comparisons}
\frac{g'(y)}{g'(x)}\le\frac{w(y)}{w(x)}\le\frac{K_d+K_c}{K_d-K_c},
\end{equation}
where $K_c$ and $K_d$ are the derivatives of $K$ with respect to $c$ and $d$, evaluated at $(c,d(c))$, and $w(u)=\sqrt{\tanh(u/2)/u}$ for $u\neq 0$ and $w(0)=1/\sqrt{2}$. Lemmas~\ref{lem:quantile-ratio} and~\ref{lem:hyperbolic-integral} give the first and second inequalities, respectively. Note that the weight $w$ is smooth, positive, and even, and the expansion $\tanh(u/2)/u=\frac{1}{2}-\frac{u^2}{24}+O(u^4)$ explains its continuous value at zero. 

\begin{figure}[t]
    \centering
    \includegraphics[scale=0.89]{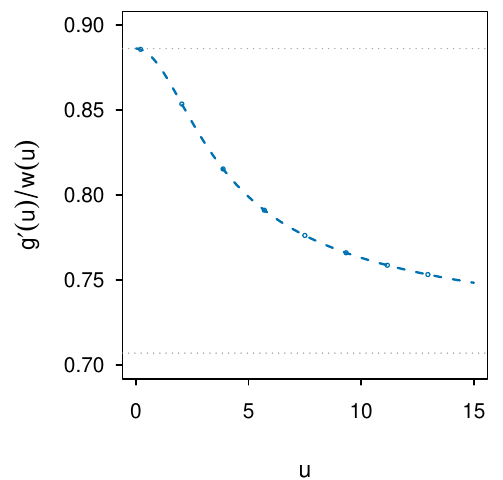}
    \caption{Ratio of $g'$ to $w$ demonstrates the inequality of Lemma~\ref{lem:quantile-ratio}.}
    \label{fig:gprime_over_w}
\end{figure}

\begin{restatable}[Gaussian derivative comparison]{lemma}{quantileratio}\label{lem:quantile-ratio}
The function $u\mapsto g'(u)/w(u)$ is nonincreasing on $[0,\infty)$. Consequently, for $y>|x|$, $g'(y)/g'(x)\le w(y)/w(x)$.
\end{restatable}

\begin{proof}[Proof of Lemma~\ref{lem:quantile-ratio}]
We aim to show that $r:=g'/w$ satisfies $r'\le 0$ by reducing this to the nonnegativity of a function $\Lambda$ and ruling out a negative interior minimum of $\Lambda$.

All derivatives in this proof are with respect to $u$. Let $s(u):=\tanh(u/2)$ and $\kappa(u):=-w'(u)/w(u)$. Recall that the function $w$ is smooth, positive, and even, so $\kappa(0)=0$. Differentiating $\Phi(g)=\sigma$ gives $\varphi(g)g'=\sigma'>0$, so $r>0$.

Since $r>0$, it suffices to determine the sign of $r'/r=g''/g'-w'/w$. For $u>0$, $w^2=s/u$ gives $w'/w=\frac{1}{2}(s'/s-1/u)$, and $s'=\frac{1}{2}\operatorname{sech}^2(u/2)$ gives $s'/s=1/\sinh u$. Consequently, $\kappa=\frac{1}{2}(1/u-1/\sinh u)$ and $\kappa'=-1/(2u^2)+\cosh u/(2\sinh^2 u)$. Taking logarithms of $\varphi(g)g'=\sigma'$ and differentiating gives $-gg'+g''/g'=\sigma''/\sigma'=1-2\sigma=-s$. Therefore
$$
r'/r=gg'-s+\kappa=-N,
$$
where $N:=s-\kappa-gg'$, and it suffices to show $N\ge 0$ on $(0,\infty)$.

Since $g$ has no closed form and is complicated to deal with, we study $N$ through a differential equation instead of bounding it directly. Differentiating and using $g''=g'(gg'-s)$ gives $N'=s'-\kappa'-(g')^2-gg'(gg'-s)$. Expansion gives $(\kappa-gg')N-gg'(gg'-s)=\kappa(s-\kappa)$, so the terms containing $gg'$ combine with $N$ into
\begin{equation}\label{eq:N-linear}
N'+(\kappa-gg')N=s'-\kappa'+\kappa(s-\kappa)-(g')^2.
\end{equation}
This is a linear first-order equation in $N$. We therefore consider the usual integrating factor $A>0$ with $A'/A=\kappa-gg'$, which turns its left side into $(AN)'/A$. Since $\kappa=-(\log w)'$ and $-gg'=(\log\varphi(g))'$, we take $A:=\varphi(g)/w$ and define
\begin{equation}\label{eq:Lambda-def}
\Lambda:=AN=\frac{\varphi(g)}{w}N=\frac{\varphi(g)(s-\kappa)-g\sigma'}{w},
\end{equation}
where the last equality uses $\varphi(g)g'=\sigma'$. Since $A>0$, $\Lambda$ has the same sign as $N$. Multiplying \eqref{eq:N-linear} by $A$, and writing $(g')^2=w^2r^2$ and $C:=(s'-\kappa'+\kappa(s-\kappa))/w^2$, gives
\begin{equation}\label{eq:Lambda-ode}
\Lambda'=\varphi(g)w(C-r^2).
\end{equation}

\begin{figure}[t]
    \centering
    \includegraphics[scale=0.89]{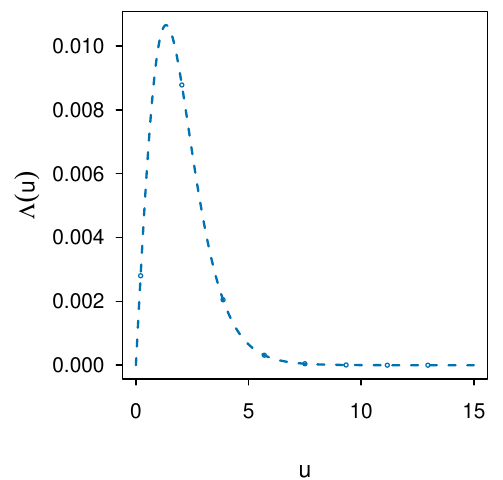}
    \includegraphics[scale=0.89]{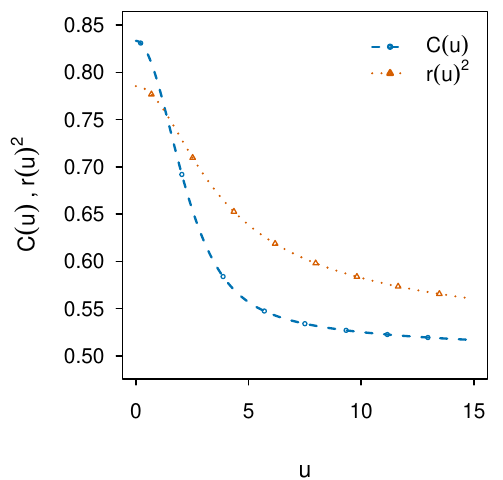}
    \caption{The trends of functions $\Lambda,\,C,\,r^2$ used in Lemma~\ref{lem:quantile-ratio}.}
    \label{fig:Lambda_C_rsquare}
\end{figure}

Since $\Lambda$ is smooth and $s(0)=\kappa(0)=g(0)=0$, \eqref{eq:Lambda-def} gives $\Lambda(0)=0$. For $u>0$, the identity $\coth u-1/\sinh u=s$ gives $s-\kappa=\frac{1}{2}(s+\coth u-1/u)>0$ (since $s>0$, and $\coth u>1/u$ because $u\cosh u-\sinh u$ equals zero at $u=0$ and has derivative $u\sinh u>0$). Dropping the positive term in the last expression of \eqref{eq:Lambda-def} gives $\Lambda\ge-E$ with $E:=g\sigma'/w$. Since $1-\Phi(g(u))=1-\sigma(u)=1/(1+e^u)$, the Chernoff bound $1-\Phi(x)\le e^{-x^2/2}$ at $x=g(u)>0$ gives $1/(1+e^u)\le e^{-g(u)^2/2}$, so $g(u)\le\sqrt{2\log(1+e^u)}\le\sqrt{2(u+\log 2)}$. Also $\sigma'(u)\le e^{-u}$, and $1/w(u)\le\sqrt{u/\tanh(1/2)}$ for $u\ge 1$. Hence $0\le E(u)\le e^{-u}\sqrt{2u(u+\log 2)/\tanh(1/2)}\to 0$, so $\liminf_{u\to\infty}\Lambda(u)\ge 0$.

Now, we aim to prove that \emph{$C$ is strictly decreasing on $(0,\infty)$.} We write $C'$ as a negative factor times a function $H$ and show $H>0$ from its power series. Substituting $s'=s/\sinh u$ and $s=(\cosh u-1)/\sinh u$,
\begin{align*}
s'-\kappa'+(s-\kappa)\kappa&=\frac{s}{2\sinh u}+\frac{s}{2u}+\frac{1}{4u^2}+\frac{1}{2u\sinh u}-\frac{2\cosh u+1}{4\sinh^2 u}\\
&=\frac{1}{4u^2}+\frac{\cosh u}{2u\sinh u}-\frac{3}{4\sinh^2 u}.
\end{align*}
Since $1/w^2=u\coth(u/2)$, $C(u)=\coth(u/2)B(u)/4$ with $B(u):=1/u+2\coth u-3u/\sinh^2 u$. Then $B'=-1/u^2-5/\sinh^2 u+6u\cosh u/\sinh^3 u$ and $\frac{d}{du}\coth(u/2)=-\coth(u/2)/\sinh u$, so the product rule gives 
$$
C'(u)=\frac{\coth(u/2)}{4}\left(B'-\frac{B}{\sinh u}\right)=-\frac{\coth(u/2)}{4u^2\sinh^3 u}H(u),
$$ 
where $H(u)=\sinh^3 u+u\sinh^2 u+u^2(5+2\cosh u)\sinh u-3u^3(2\cosh u+1)$. The factor multiplying $-H(u)$ is positive for $u>0$, so it remains to prove $H(u)>0$. Using $\sinh^3 u=(\sinh(3u)-3\sinh u)/4$, $\sinh^2 u=(\cosh(2u)-1)/2$, and $2\cosh u\sinh u=\sinh(2u)$,
\begin{align*}
H(u)={}&\frac{\sinh(3u)-3\sinh u}{4}+\frac{u}{2}(\cosh(2u)-1)+u^2(5\sinh u+\sinh(2u))-6u^3\cosh u-3u^3.
\end{align*}
Each term of $H$ is odd in $u$, so $H$ is odd and its power series contains only the powers $u^{2n+1}$. The coefficients of $u$, $u^3$, and $u^5$ equal zero because $\frac34-\frac34=0$, $1+1+7-6-3=0$, and $\frac{1}{2}+\frac{1}{3}+\frac{13}{6}-3=0$. For $n\ge2$, the coefficient of $u^{2n+1}/(2n+1)!$ is $3^{2n+1}/4$ from $\frac14\sinh(3u)$, $-3/4$ from $-\frac34\sinh u$, $(2n+1)2^{2n-1}$ from $\frac u2\cosh(2u)$, $2n(2n+1)(5+2^{2n-1})$ from $u^2(5\sinh u+\sinh(2u))$, and $-6(2n-1)(2n)(2n+1)$ from $-6u^3\cosh u$, while $-u/2$ and $-3u^3$ affect only $u$ and $u^3$. Hence $H(u)=\sum_{n=3}^{\infty}A_n\frac{u^{2n+1}}{(2n+1)!}$, where
\begin{align*}
A_n&=\frac{3^{2n+1}-3}{4}+(2n+1)2^{2n-1}+2n(2n+1)(5+2^{2n-1})-6(2n-1)(2n)(2n+1)\\
&=\frac{3^{2n+1}-3}{4}+(2n+1)^2 2^{2n-1}-2n(2n+1)(12n-11).
\end{align*}
For $n\ge 3$, $2^{2n-1}\ge 12n-11$ by induction, since $2^5=32\ge 25$ and $2^{2n+1}=4\cdot 2^{2n-1}\ge 48n-44\ge 12(n+1)-11$. Hence $(2n+1)^2 2^{2n-1}>2n(2n+1)(12n-11)$, and with $(3^{2n+1}-3)/4>0$ this gives $A_n>0$. Every term of the series for $H(u)$ is therefore positive for $u>0$, so $H(u)>0$ and $C'(u)<0$.

The facts $\Lambda(0)=0$, $\liminf_{u\to\infty}\Lambda(u)\ge0$, and $C'<0$ now combine to show that $\Lambda$ cannot take a negative value. Suppose $\Lambda(u_0)<0$ for some $u_0>0$. Since $\liminf_{u\to\infty}\Lambda(u)\ge 0$, there is $R>u_0$ with $\Lambda(R)>\Lambda(u_0)$. Since $\Lambda(0)=0>\Lambda(u_0)$, the minimum of $\Lambda$ on $[0,R]$ is attained at an interior point $v$, where $\Lambda(v)<0$, $\Lambda'(v)=0$, and $\Lambda''(v)\ge 0$. First, $\Lambda(v)<0$ gives $N(v)<0$, so $r'(v)=-r(v)N(v)>0$. Second, $\Lambda'(v)=0$ and \eqref{eq:Lambda-ode} give $C(v)=r(v)^2$. Third, differentiating \eqref{eq:Lambda-ode} gives $\Lambda''=(\varphi(g)w)'(C-r^2)+\varphi(g)w(C'-2rr')$, whose first term equals zero at $v$. Hence $\Lambda''(v)=\varphi(g(v))w(v)[C'(v)-2r(v)r'(v)]<0$, because $C'(v)<0$, $r(v)>0$, and $r'(v)>0$. This contradicts $\Lambda''(v)\ge 0$.

This contradiction shows that $\Lambda\ge 0$ on $(0,\infty)$. Since $A>0$ and $r>0$, it follows that $N=\Lambda/A\ge 0$ and hence $r'=-rN\le 0$ on $(0,\infty)$. Since $r$ is continuous on $[0,\infty)$, the function $r=g'/w$ is therefore nonincreasing on $[0,\infty)$. Finally, let $y>|x|$. Since $0\le|x|<y$ and $g'$ and $w$ are even, $g'(y)/w(y)\le g'(|x|)/w(|x|)=g'(x)/w(x)$, which rearranges to $g'(y)/g'(x)\le w(y)/w(x)$ because $g'$ and $w$ are positive.
\end{proof}

\begin{figure}[t]
    \centering
    \includegraphics[scale=0.89]{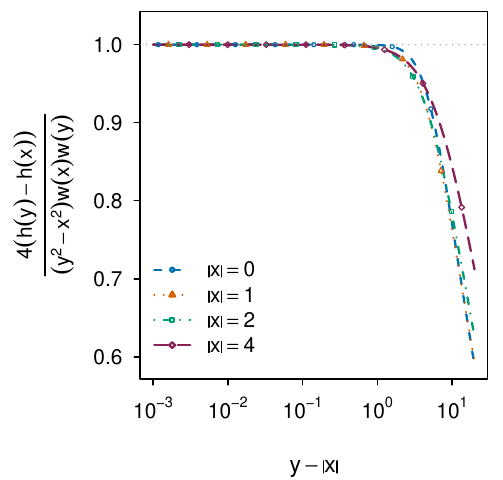}
    \caption{Illustration of the inequality of Lemma~\ref{lem:hyperbolic-integral}.}
    \label{fig:hyperbolic}
\end{figure}

\begin{restatable}[Hyperbolic integral comparison]{lemma}{hyperbolicintegral}\label{lem:hyperbolic-integral}
For $y>|x|$,
\begin{equation}\label{eq:hyperbolic-target}
4[h(y)-h(x)]\le(y^2-x^2)w(x)w(y).
\end{equation}
\end{restatable}

\begin{proof}[Proof of Lemma~\ref{lem:hyperbolic-integral}]
Define
\[
W(t)=\begin{cases}\tanh(\sqrt t/2)/\sqrt t,&t>0,\\1/2,&t=0,\end{cases}
\]
so that $W(u^2)=w(u)^2$ and $\frac{d}{dt}h(\sqrt t)=\frac14W(t)$. Since $h$ is even, $4[h(y)-h(x)]=\int_{x^2}^{y^2}W(t)\,dt$, and \eqref{eq:hyperbolic-target} becomes
\begin{equation}\label{eq:W-average-target}
\frac{1}{y^2-x^2}\int_{x^2}^{y^2}W(t)\,dt\le\sqrt{W(x^2)W(y^2)}.
\end{equation}
To write $W$ as a sum of simple terms, take logarithms in the Euler product $\cosh z=\prod_{n=1}^{\infty}\bigl(1+\frac{4z^2}{\pi^2(2n-1)^2}\bigr)$ \cite{dlmf-cosh} at $z=\sqrt t/2$. With $\lambda_n=\pi^2(2n-1)^2$, this gives $\log\cosh(\sqrt t/2)=\sum_{n=1}^{\infty}\log(1+t/\lambda_n)$, and differentiating in $t$ gives
\begin{equation}\label{eq:W-series}
W(t)=4\sum_{n=1}^{\infty}f_n(t),
\end{equation}
where $f_n(t):=\frac{1}{t+\lambda_n}$. Each $f_n$ satisfies \eqref{eq:W-average-target} with $W$ replaced by $f_n$. With $a_n=x^2+\lambda_n$ and $b_n=y^2+\lambda_n$, the average of $f_n$ over $[x^2,y^2]$ is $(\log b_n-\log a_n)/(b_n-a_n)$, the reciprocal of the logarithmic mean of $a_n$ and $b_n$. Since the logarithmic mean is at least the geometric mean $\sqrt{a_nb_n}$ \cite{carlson1972logarithmic},
\begin{equation}\label{eq:reciprocal-average}
\frac{1}{y^2-x^2}\int_{x^2}^{y^2}f_n(t)\,dt\le\frac{1}{\sqrt{a_nb_n}}=\sqrt{f_n(x^2)f_n(y^2)}.
\end{equation}

Integrating the positive series \eqref{eq:W-series} term by term, then applying \eqref{eq:reciprocal-average} and Cauchy--Schwarz,
\begin{align*}
\frac{1}{y^2-x^2}\int_{x^2}^{y^2}W(t)\,dt
\le4\sum_{n=1}^{\infty}\sqrt{f_n(x^2)}\sqrt{f_n(y^2)}
\le4\sqrt{\Bigl(\sum_{n=1}^{\infty}f_n(x^2)\Bigr)\Bigl(\sum_{n=1}^{\infty}f_n(y^2)\Bigr)}.
\end{align*}
This proves \eqref{eq:W-average-target}, and hence \eqref{eq:hyperbolic-target} follows.
\end{proof}

\begin{figure}[t]
    \centering
    \includegraphics[scale=0.89]{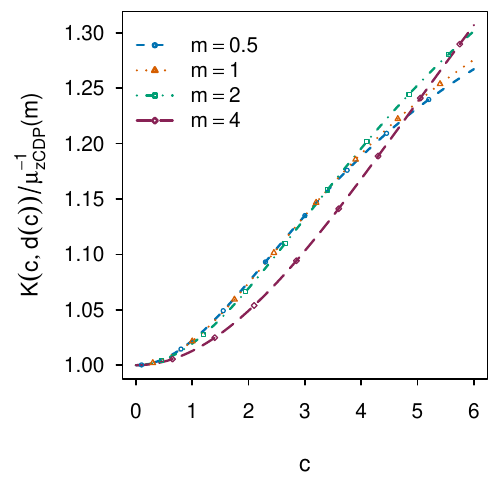}
    \caption{Ratio of the selected-order bound to the zCDP budget of randomized response, plotted against the midpoint $c$ of the log-odds of the two Bernoulli distributions, for Bernoulli pairs with a fixed GDP parameter. The ratio is at least one and equals one at $c = 0$, which corresponds to $\mathrm{RR}_t$.}
    \label{fig:Kapp}
\end{figure}

We now combine Lemmas~\ref{lem:quantile-ratio} and~\ref{lem:hyperbolic-integral}. Along a curve of fixed Gaussian separation, they imply that the lower bound $K$ in \eqref{eq:K-lower-bound} cannot decrease as the midpoint moves away from zero (Figure~\ref{fig:Kapp}).

\begin{lemma}[Symmetrization at fixed Gaussian separation]\label{lem:symmetrization}
Let $c\ge 0$, $d>0$, and $a>0$ satisfy $g(c+d)-g(c-d)=2g(a)$. Then $K(c,d)\ge K(0,a)=a\tanh(a/2)$.
\end{lemma}

\begin{proof}[Proof of Lemma~\ref{lem:symmetrization}]
We show that $K$ does not decrease along the curve of Gaussian separation $2g(a)$ as $c$ moves away from zero. For each $c\ge 0$, the function $d\mapsto g(c+d)-g(c-d)$ increases strictly from zero at $d=0$ to infinity because its derivative is $g'(c+d)+g'(c-d)>0$. So, there exists a unique $d(c)>0$ that satisfies $g(c+d(c))-g(c-d(c))=2g(a)$, and $d(0)=a$ because $g$ is odd. By the implicit function theorem and uniqueness, $d(\cdot)$ is smooth, so $d(c)\to a$ as $c\downarrow0$. Since $K$ is continuous at $c=0$ (Appendix~\ref{app:log-odds-selected-order}), the map $c\mapsto K(c,d(c))$ is continuous at zero with value $K(0,a)$, and it suffices to show $\frac{d}{dc}K(c,d(c))\ge 0$ for $c>0$.

Given $c>0$ and write $d=d(c)$, $x=c-d$, $y=c+d$, $L=h(y)-h(x)$, and $B=y^2-x^2=4cd$. Since $y-x=2d>0$ and $y+x=2c>0$, we have $y>|x|$, so $L>0$ and $B>0$. Differentiating the constraint $g(c+d(c))-g(c-d(c))=2g(a)$ gives $g'(y)(1+d'(c))-g'(x)(1-d'(c))=0$, so $d'(c)=(g'(x)-g'(y))/(g'(x)+g'(y))$. By the chain rule,
\begin{equation}\label{eq:fixed-separation-K-derivative}
\frac{d}{dc}K(c,d(c))=K_c+K_d d'(c)=\frac{g'(x)(K_d+K_c)-g'(y)(K_d-K_c)}{g'(x)+g'(y)}.
\end{equation}
Differentiating $K(c,d)=d[h(c+d)-h(c-d)]/c$ gives $K_c=-dL/c^2+d[h'(y)-h'(x)]/c$ and $K_d=L/c+d[h'(y)+h'(x)]/c$. Adding and subtracting, with $B=4cd$, yields $K_d+K_c=N/(2c^2)$ and $K_d-K_c=D/(2c^2)$, where $N:=2xL+Bh'(y)$ and $D:=2yL+Bh'(x)$.

The sign of \eqref{eq:fixed-separation-K-derivative} can be read off once we know $D>0$. If $x\ge 0$, then $2yL>0$ and $Bh'(x)\ge 0$. If $x=-t<0$, then $0<t<y$, and evenness and convexity of $h$ give $L=h(y)-h(t)\ge(y-t)h'(t)$. Since $h'(x)=-h'(t)$, it follows that $D=2yL-(y^2-t^2)h'(t)\ge[2y(y-t)-(y^2-t^2)]h'(t)=(y-t)^2 h'(t)>0$. Thus $D>0$ in both cases, so $K_d-K_c>0$. Since also $g'(x)>0$, \eqref{eq:fixed-separation-K-derivative} shows that $\frac{d}{dc}K(c,d(c))\ge 0$ if and only if $g'(y)/g'(x)\le(K_d+K_c)/(K_d-K_c)=N/D$.

We verify this condition through the intermediate ratio $w(y)/w(x)$, as in \eqref{eq:two-comparisons}. Since $y>|x|$, Lemma~\ref{lem:quantile-ratio} yields $g'(y)/g'(x)\le w(y)/w(x)$. To compare $w(y)/w(x)$ with $N/D$, use $h'(u)=uw(u)^2/2$ to calculate
\begin{align*}
Nw(x)-Dw(y)
&=2L[xw(x)-yw(y)]+B[h'(y)w(x)-h'(x)w(y)]\\
&=2L[xw(x)-yw(y)]+\tfrac{B}{2}w(x)w(y)[yw(y)-xw(x)]\\
&=[yw(y)-xw(x)]\left[\tfrac{B}{2}w(x)w(y)-2L\right].
\end{align*}
The first factor is positive because $y>|x|$, and $uw(u)$ is odd and equals $\sqrt{u\tanh(u/2)}$ for $u>0$, which is strictly increasing because $u$ and $\tanh(u/2)$ are. The second factor is nonnegative by Lemma~\ref{lem:hyperbolic-integral}, since $4L\le Bw(x)w(y)$. Hence $Nw(x)\ge Dw(y)$, and dividing by $Dw(x)>0$ yields $w(y)/w(x)\le N/D$. Combining the two comparisons gives $g'(y)/g'(x)\le N/D$, so $\frac{d}{dc}K(c,d(c))\ge 0$.

Thus, $c\mapsto K(c,d(c))$ is nondecreasing on $(0,\infty)$, and by continuity at zero, $K(c,d(c))\ge K(0,a)=a\tanh(a/2)$ for every $c\ge 0$. Since uniqueness gives $d=d(c)$ for the pair $(c,d)$ in the lemma, this proves Lemma~\ref{lem:symmetrization}.
\end{proof}

\subsection{Completing the zCDP Bernoulli bound and conversion}\label{app:zcdp-completion}
Equation~\eqref{eq:K-lower-bound} and Lemma~\ref{lem:symmetrization} now give the zCDP comparison between the divergence budget $K_{\mathrm{zCDP}}$ in \eqref{eq:H-definition} and the inverse conversion $\mu_{\mathrm{zCDP}}^{-1}$ in \eqref{eq:R-definition}. The symmetric case follows by taking the R\'enyi order to one.

\begin{restatable}[The zCDP Bernoulli comparison]{lemma}{bernoullibound}\label{lem:bernoulli-bound}
For every $0<p<q<1$,
\begin{equation}\label{eq:bernoulli-target}
K_{\mathrm{zCDP}}(p,q)\ge\mu_{\mathrm{zCDP}}^{-1}(m_{p,q}).
\end{equation}
Equality holds for the symmetric pair $p=\sigma(-a)$, $q=\sigma(a)$ with Gaussian separation $m_{p,q}=2\Phi^{-1}(\sigma(a))$.
\end{restatable}

\begin{proof}[Proof of Lemma~\ref{lem:bernoulli-bound}]
By the reflection in Appendix~\ref{app:log-odds-selected-order}, we may assume $c\ge0$, with $p=\sigma(c-d)$, $q=\sigma(c+d)$, and $d>0$. We first show $K_{\mathrm{zCDP}}(p,q)\ge K(c,d)$. For $c>0$, this is \eqref{eq:K-lower-bound}. For $c=0$, letting $\eta\downarrow1$ in \eqref{eq:H-definition} gives $K_{\mathrm{zCDP}}(p,q)\ge\KL(\Bern(p)\|\Bern(q))=d\tanh(d/2)=K(0,d)$. Now let $a>0$ satisfy $2g(a)=m_{p,q}$, as in Appendix~\ref{app:symmetrization}. Lemma~\ref{lem:symmetrization} and \eqref{eq:R-symmetric} give
\[
K_{\mathrm{zCDP}}(p,q)\geq K(c,d)\geq K(0,a)=\mu_{\mathrm{zCDP}}^{-1}(2g(a))=\mu_{\mathrm{zCDP}}^{-1}(m_{p,q}).
\]
For $p=\sigma(-a)$ and $q=\sigma(a)$, Proposition~\ref{prop:rr} gives $K_{\mathrm{zCDP}}(p,q)=a\tanh(a/2)$, so equality holds.
\end{proof}

\sharpconversion*

\begin{proof}[Proof of Theorem~\ref{thm:sharp-gdp-conversions}]
For $\rho=0$, the divergence bound forces neighboring output distributions to coincide, giving $0$-GDP. Given any $\rho>0$, suppose a $\rho$-zCDP mechanism fails $\mu_{\mathrm{zCDP}}(\rho)$-GDP, and let $p<q$ be the type I error and power of the test from Appendix~\ref{app:bernoulli-reduction}. If $p=0$ or $q=1$, a directed R\'enyi divergence of the binary pair is infinite. Otherwise, as $0<p<q<1$ and $m_{p,q}>\mu_{\mathrm{zCDP}}(\rho)$, Lemma~\ref{lem:bernoulli-bound} and strict monotonicity of $\mu_{\mathrm{zCDP}}^{-1}$ give
\[
K_{\mathrm{zCDP}}(p,q)\ge\mu_{\mathrm{zCDP}}^{-1}(m_{p,q})
>\mu_{\mathrm{zCDP}}^{-1}(\mu_{\mathrm{zCDP}}(\rho))=\rho,
\]
and hence, by \eqref{eq:H-definition}, there exist an order $\eta>1$ and a direction that gives a binary divergence greater than $\rho\eta$. In either case, \eqref{eq:data-processing} transfers the violation to the original neighboring distributions, which contradicts $\rho$-zCDP because the neighboring relation is symmetric.

By Proposition~\ref{prop:rr}, randomized response with parameter $t_\rho$ is exactly $\rho$-zCDP and exactly $\mu_{\mathrm{zCDP}}(\rho)$-GDP, so no smaller universal parameter is possible.
\end{proof}

\section{Proof of the sharp bounded-range-to-GDP conversion}\label{app:bounded-range-conversion-proof}

The proof follows the contrapositive of Appendix~\ref{app:zcdp-conversion-proof}, with privacy-loss width in place of R\'enyi divergence. The only analytic ingredient reused is Lemma~\ref{lem:quantile-ratio}.

With $a$ as in \eqref{eq:inverse-common-parameter}, $\mu_{\mathrm{BR}}(2a)=m$, so the functional inverse $m\ge0$ is
\begin{equation}\label{eq:br-inverse}
\mu_{\mathrm{BR}}^{-1}(m)=2\log\frac{\Phi(m/2)}{\Phi(-m/2)}.
\end{equation}
It is strictly increasing, since the numerator increases and the positive denominator decreases with $m$.

Suppose an $\varepsilon$-BR mechanism fails $\mu_{\mathrm{BR}}(\varepsilon)$-GDP. The construction of Appendix~\ref{app:bernoulli-reduction} with $\mu=\mu_{\mathrm{BR}}(\varepsilon)$ gives a test $\phi$ and a binary pair $\Bern(p),\Bern(q)$ satisfying $q>\Phi(\Phi^{-1}(p)+\mu_{\mathrm{BR}}(\varepsilon))$, so $p<q$ and
\begin{equation*}
m_{p,q}=\Phi^{-1}(q)-\Phi^{-1}(p)>\mu_{\mathrm{BR}}(\varepsilon).
\end{equation*}

The binary pair also preserves bounded range. Since $P$ and $Q$ are mutually absolutely continuous, $p=0$ would force $q=0$ and $q=1$ would force $p=1$, so $0<p<q<1$. The bounded-range assumption provides $L=dQ/dP$ and constants $0<A\le B<\infty$ such that $A\le L\le B$ and $\log(B/A)\le\varepsilon$. The binary likelihood ratios satisfy
\[
\frac qp=\frac{\mathbb E_P[L\phi]}{\mathbb E_P\phi}\in[A,B],\qquad
\frac{1-q}{1-p}=\frac{\mathbb E_P[L(1-\phi)]}{\mathbb E_P[1-\phi]}\in[A,B],
\]
since both are weighted averages of $L$. Since $p<q$, the binary width satisfies
\[
K_{\mathrm{BR}}(p,q)=\log\frac qp-\log\frac{1-q}{1-p}\le\log(B/A)\le\varepsilon.
\]

For an $\varepsilon$-logistic DP mechanism, the same test instead satisfies $1-q\ge L_\varepsilon(p)$. Since $L_\varepsilon(0)=1$ and $L_\varepsilon(p)>0$ for $p<1$, a pair with $p<q$ is again interior. Rearrangement gives $q(1-p)\le e^\varepsilon p(1-q)$ and $K_{\mathrm{BR}}(p,q)\le\varepsilon$.

In the coordinates of Appendix~\ref{app:log-odds-selected-order}, $p=\sigma(c-d)$ and $q=\sigma(c+d)$, so
\begin{equation*}
K_{\mathrm{BR}}(p,q)=\log\frac{q(1-p)}{p(1-q)}=2d,
\end{equation*}
and the separation is $m_{p,q}=g(c+d)-g(c-d)$. Since the width depends only on $d$, we maximize the separation over $c$ at fixed $d$.

Lemma~\ref{lem:quantile-ratio} gives a positive nonincreasing ratio $g'/w$, and $w$ is strictly decreasing on $[0,\infty)$ because $w'(u)/w(u)=\frac12(1/\sinh u-1/u)<0$ for $u>0$. Hence $g'=(g'/w)w$ is positive and strictly decreasing on $[0,\infty)$, and it is even because $g$ is odd. Oddness of $g$ also makes the separation even in $c$, and for $c>0$ its derivative $g'(c+d)-g'(|c-d|)$ is negative because $c+d>|c-d|$. The separation is therefore maximized at $c=0$, which gives
\begin{equation*}
m_{p,q}=g(c+d)-g(c-d)\le2g(d).
\end{equation*}

Take $a$ from \eqref{eq:inverse-common-parameter} with $m=m_{p,q}$, so that $2g(a)=m_{p,q}\le 2g(d)$. Strict monotonicity of $g$ gives $a\le d$, and by \eqref{eq:br-inverse},
\begin{equation}\label{eq:br-binary-bound}
K_{\mathrm{BR}}(p,q)=2d\ge2a=\mu_{\mathrm{BR}}^{-1}(m_{p,q}).
\end{equation}

\brsharpconversion*

\begin{proof}[Proof of Theorem~\ref{thm:br-gdp-conversion}]
For $\varepsilon>0$, suppose an $\varepsilon$-BR or $\varepsilon$-logistic DP mechanism fails $\mu_{\mathrm{BR}}(\varepsilon)$-GDP. Then,  $K_{\mathrm{BR}}(p,q)\le\varepsilon$ and $m_{p,q}>\mu_{\mathrm{BR}}(\varepsilon)$. By \eqref{eq:br-binary-bound} and strict monotonicity of $\mu_{\mathrm{BR}}^{-1}$,
\[
\varepsilon\ge K_{\mathrm{BR}}(p,q)
\ge\mu_{\mathrm{BR}}^{-1}(m_{p,q})
>\mu_{\mathrm{BR}}^{-1}(\mu_{\mathrm{BR}}(\varepsilon))
=\varepsilon,
\]
a contradiction. For $\varepsilon=0$, bounded range makes the likelihood ratio constant and normalization makes it one; logistic DP gives $T(P,Q)(\alpha)\ge1-\alpha$. In either case neighboring distributions coincide and satisfy $0$-GDP.

For sharpness, randomized response with parameter $\varepsilon/2$ has privacy-loss values $\pm\varepsilon/2$, so it is exactly $\varepsilon$-BR and hence $\varepsilon$-logistic DP. By Proposition~\ref{prop:rr}, its exact GDP parameter is $2\Phi^{-1}(\sigma(\varepsilon/2))=\mu_{\mathrm{BR}}(\varepsilon)$, so no smaller parameter is valid for either class.
\end{proof}

\section{Proofs of conversion properties and privacy accounting}\label{app:conversion-properties-accounting}
Our main question here is whether to convert each component before composition or to compose the zCDP or logistic DP budgets before conversion. We first establish a Gaussian integral identity that makes the inverse-conversion derivatives tractable.

\begin{lemma}[Gaussian integral identity]\label{lem:gauss_integral_identity}
Let $\Theta\sim\operatorname{Unif}[0,\pi/4]$ and define $A(\Theta)=1/(8\cos^2\Theta)$. For every $s\geq0$,
\[
4\Phi(\sqrt{s}/2)\Phi(-\sqrt{s}/2)
=\frac4\pi\int_0^{\pi/4} \exp\left(-\frac{s}{8\cos^2\theta}\right)\,d\theta
=\mathbb E[e^{-sA}].
\]
\end{lemma}
\begin{proof}
Put $t=\sqrt{s}/2$, and let $Z_1,Z_2$ be independent standard normal variables. Independence gives $\Pr(\max(|Z_1|,|Z_2|)>t)=1-(2\Phi(t)-1)^2=4\Phi(t)\Phi(-t)$. In polar coordinates, a point in the sector $0\le\theta\le\pi/4$ lies outside $[-t,t]^2$ precisely when $r>t/\cos\theta$. The eight sectors have equal probability, so
\begin{align*}
\Pr(\max(|Z_1|,|Z_2|)>t)
&=\frac{8}{2\pi}\int_0^{\pi/4}\int_{t/\cos\theta}^{\infty}e^{-r^2/2}r\,dr\,d\theta
=\frac{4}{\pi}\int_0^{\pi/4}\exp\!\left(-\frac{t^2}{2\cos^2\theta}\right)d\theta.
\end{align*}
Substituting $t^2=s/4$, the expectation representation follows because $\Theta$
has density $4/\pi$ on $[0,\pi/4]$.
\end{proof}

Lemma~\ref{lem:gauss_integral_identity} expresses the product as the Laplace transform of the random variable $A=(8\cos^2\Theta)^{-1}$, where $\Theta\sim\operatorname{Unif}[0,\pi/4]$. This representation makes differentiation with respect to \(s\) particularly convenient because $\frac{d^k}{ds^k}\mathbb E[e^{-sA}] =(-1)^k\mathbb E[A^k e^{-sA}]$.

\conversionconvexity*

\begin{figure}[t]
    \centering
    \includegraphics[scale=0.89]{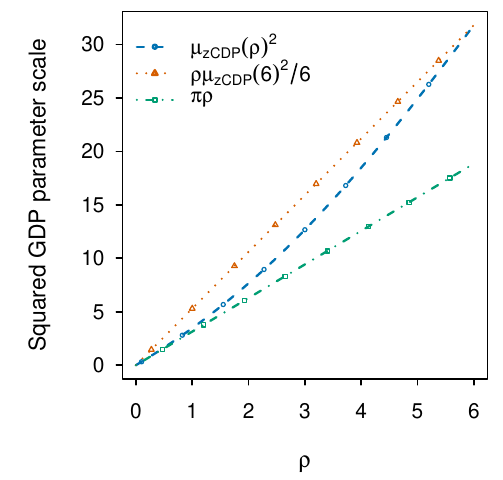}
    \includegraphics[scale=0.89]{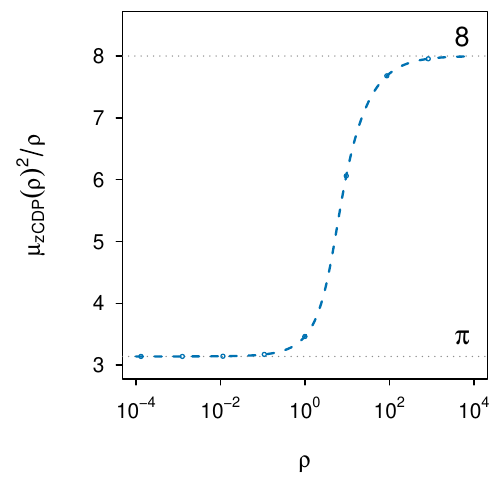}
    \caption{Shape of the sharp conversion from zCDP to GDP. Left: $\rho\mapsto\mu_{\mathrm{zCDP}}(\rho)^2$ is strictly convex. Right: $\mu_{\mathrm{zCDP}}(\rho)^2/\rho$ increases from $\pi$ to $8$, the sharp limiting constants in the conversion bounds that hold for every $\rho$-zCDP mechanism.}
    \label{fig:muzCDP2}
\end{figure}

\begin{proof}[Proof of Theorem~\ref{thm:conversion-shape}]
Write $F_{\mathrm{zCDP}}(\rho)=\mu_{\mathrm{zCDP}}(\rho)^2$, with $F_{\mathrm{zCDP}}(0)=0$. Since this expression involves the inverse Gaussian CDF, we study its inverse to work instead with the Gaussian CDF, whose derivative is explicit. This also avoids implicitly differentiating $\rho=t_\rho\tanh(t_\rho/2)$ to obtain $t_\rho'(\rho)$. It suffices to show that $F_{\mathrm{zCDP}}^{-1}$ is strictly increasing and strictly concave, which we do by proving that its derivative is positive and strictly decreasing.

To express the inverse, take $s=\mu_{\mathrm{zCDP}}(\rho)^2$ as the input, so $F_{\mathrm{zCDP}}^{-1}(s)=\mu_{\mathrm{zCDP}}^{-1}(\sqrt{s})$ is the corresponding zCDP budget $\rho$. For squared Gaussian separation $s\ge0$, let $t(s)$ denote the corresponding randomized-response parameter. The identity $\sqrt{s}=2\Phi^{-1}(\sigma(t))$ gives $\Phi(\sqrt{s}/2)=\sigma(t)$. Taking logistic odds and then logarithms yields $t(s)=\log\frac{\Phi(\sqrt{s}/2)}{\Phi(-\sqrt{s}/2)}$ and $ F_{\mathrm{zCDP}}^{-1}(s)=t(s)\tanh(t(s)/2)$. As $s$ increases, the numerator in the odds increases and the denominator decreases, so $t(s)$ is strictly increasing. Notice that $t(\mu_{\mathrm{zCDP}}(\rho)^2)=t_\rho$; this is the same randomized-response parameter, now expressed as a function of the squared GDP parameter.

\textit{The derivative of the inverse conversion.}
We want an expression for $(F_{\mathrm{zCDP}}^{-1})'(s)$ whose positivity and monotonicity can be studied separately. For $s>0$, differentiating gives
\begin{align*}
(F_{\mathrm{zCDP}}^{-1})'(s)
=\tanh(t(s)/2)t'(s)+\frac{t(s)}2\operatorname{sech}^2(t(s)/2)t'(s).
\end{align*}
Factoring out $\tanh(t(s)/2)t'(s)$ leaves the factor $1+t(s)/\sinh t(s)$. This factorization follows from the double-angle identity for the hyperbolic sine. Define $J(s):=2\log\cosh(t(s)/2)$; its derivative is precisely the extracted factor. Thus
\begin{equation}\label{eq:inverse-conversion-slope}
(F_{\mathrm{zCDP}}^{-1})'(s)=\left(1+\frac{t(s)}{\sinh t(s)}\right)J'(s).
\end{equation}
We will show that both factors are positive and strictly decreasing.

The choice of $J$ also connects the derivative calculation to the Gaussian integral identity. Since $\Phi(\sqrt{s}/2)=\sigma(t(s))$ and $\Phi(-\sqrt{s}/2)=1-\sigma(t(s))$, their product satisfies $4\Phi(\sqrt{s}/2)\Phi(-\sqrt{s}/2)=1/\cosh^2(t(s)/2)$. Taking negative logarithms and applying Lemma~\ref{lem:gauss_integral_identity} gives
\[
J(s)=-\log[4\Phi(\sqrt{s}/2)\Phi(-\sqrt{s}/2)]=-\log\mathbb E[e^{-sA}],
\]
where $A=(8\cos^2\Theta)^{-1}$ and $\Theta\sim\operatorname{Unif}[0,\pi/4]$. This representation allows us to determine the signs of $J'$ and $J''$ without differentiating the Gaussian CDF product twice. Because $A$ is bounded, differentiation under the expectation is valid, and
\begin{align*}
J'(s)&=\frac{\mathbb E[Ae^{-sA}]}{\mathbb E[e^{-sA}]}=\mathbb E_s[A]>0,\\
J''(s)&=-\frac{\mathbb E[A^2e^{-sA}]}{\mathbb E[e^{-sA}]}
+\left(\frac{\mathbb E[Ae^{-sA}]}{\mathbb E[e^{-sA}]}\right)^2
=-\operatorname{Var}_s(A)<0.
\end{align*}
Here $\mathbb E_s[f(A)]:=\mathbb E[f(A)e^{-sA}]/\mathbb E[e^{-sA}]$ is expectation under the distribution obtained by weighting by $e^{-sA}$ and normalizing. The weights are strictly positive, so $A$ remains nonconstant and the variance is strictly positive. Hence $J'$ is positive and strictly decreasing.

For the other factor, $u/\sinh u$ is strictly decreasing on $(0,\infty)$, since its derivative $(\sinh u-u\cosh u)/\sinh^2u$ is negative by $u\cosh u-\sinh u>0$ (proof of Lemma~\ref{lem:quantile-ratio}). Since $t(s)$ is strictly increasing, $1+t(s)/\sinh t(s)$ is positive and strictly decreasing in $s$.

\textit{Strict convexity.}
The inverse derivative in \eqref{eq:inverse-conversion-slope} is a product of two positive, strictly decreasing functions. Therefore $F_{\mathrm{zCDP}}^{-1}$ is strictly increasing and strictly concave on $(0,\infty)$. Its continuous extension to zero is concave on $[0,\infty)$; strictness for a segment starting at zero follows by comparing average slopes of its strictly decreasing derivative. The inverse of an increasing, strictly concave function is strictly convex, so $F_{\mathrm{zCDP}}$ is strictly increasing and strictly convex on $[0,\infty)$.

\begin{figure}[t]
\centering
\includegraphics[scale=0.89]{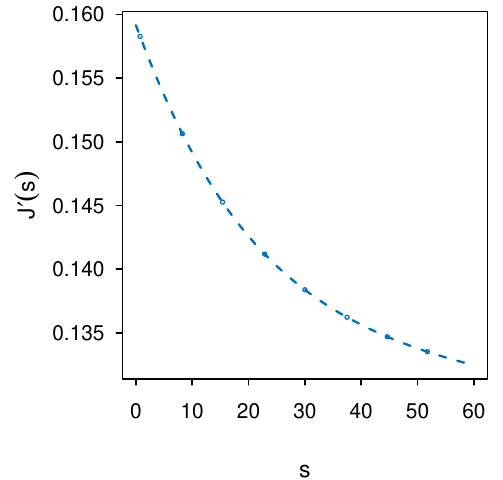}
\includegraphics[scale=0.89]{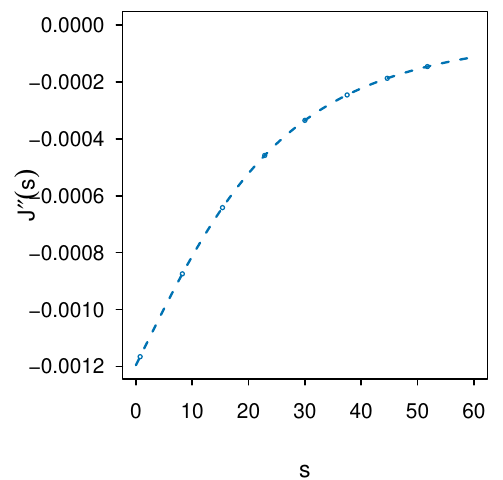}
\caption{The negative log-Laplace transform used in the zCDP inverse calculation. The signs $J'>0$ and $J''<0$ give a positive decreasing factor in \eqref{eq:inverse-conversion-slope}.}
\label{fig:J_derivatives}
\end{figure}

\textit{The two limits.}
Recall that $\rho=t_\rho\tanh(t_\rho/2)$ and $\mu_{\mathrm{zCDP}}(\rho)=2g(t_\rho)$. The map $t\mapsto t\tanh(t/2)$ is strictly increasing from zero to infinity. Consequently, $\rho\downarrow0$ corresponds to $t_\rho\downarrow0$, and $\rho\to\infty$ corresponds to $t_\rho\to\infty$. We use this parameterization to compare the numerator and denominator of $F_{\mathrm{zCDP}}(\rho)/\rho$.

\emph{As $\rho\downarrow0$.}
First, $\tanh(t/2)=t/2+o(t)$, so $\rho=t_\rho^2/2+o(t_\rho^2)$. Next, we expand $2\Phi^{-1}(\sigma(t))$ at $t=0$. The inverse function rule gives $(\Phi^{-1})'(1/2)=1/\varphi(\Phi^{-1}(1/2))=1/\varphi(0)=\sqrt{2\pi}$. Since $\sigma(0)=1/2$ and $\sigma'(0)=1/4$, the chain rule gives $2g(t)=2\left(\sqrt{2\pi}\,\frac{t}{4}\right)+o(t)
=\sqrt{\frac{\pi}{2}}\,t+o(t)$.
Consequently, $F_{\mathrm{zCDP}}(\rho)=(\pi/2)t_\rho^2+o(t_\rho^2)$. Both numerator and denominator are of order $t_\rho^2$, so dividing their expansions yields
\[
\lim_{\rho\downarrow0}\frac{F_{\mathrm{zCDP}}(\rho)}{\rho}
=\lim_{t\downarrow0}\frac{4g(t)^2}{t\tanh(t/2)}
=\frac{\pi/2}{1/2}=\pi.
\]

\emph{As $\rho\to\infty$.}
Write $z=\mu_{\mathrm{zCDP}}(\rho)/2=g(t_\rho)$. From the conversion formula, $t_\rho=\log\{\Phi(z)/\Phi(-z)\}$ and $\tanh(t_\rho/2)=2\Phi(z)-1$, so
\[
\frac{F_{\mathrm{zCDP}}(\rho)}{\rho}
=\frac{4z^2}{(2\Phi(z)-1)\log\{\Phi(z)/\Phi(-z)\}}.
\]
As $\rho\to\infty$, $z\to\infty$ and $2\Phi(z)-1\to1$. It remains to evaluate the logarithmic factor relative to $z^2$.

Both $\log\{\Phi(z)/\Phi(-z)\}$ and $z^2$ tend to infinity. The derivative ratio in L'H\^opital's rule is
\[
\frac{\varphi(z)/\Phi(z)+\varphi(z)/\Phi(-z)}{2z}
=\frac12\left[\frac{\varphi(z)}{z\Phi(z)}+\frac{\varphi(z)}{z\Phi(-z)}\right].
\]
The first term in brackets tends to zero. For the second, the Mills bounds $\frac{z}{1+z^2}\varphi(z)\le\Phi(-z)\le\varphi(z)/z$ give $z^2/(1+z^2)\le z\Phi(-z)/\varphi(z)\le1$. Taking reciprocals yields $1\le\varphi(z)/(z\Phi(-z))\le1+1/z^2$, so this term tends to one by the squeeze theorem. L'H\^opital's rule therefore gives $\lim_{z\to\infty}\log\{\Phi(z)/\Phi(-z)\}/z^2=1/2$ and $\lim_{\rho\to\infty}F_{\mathrm{zCDP}}(\rho)/\rho=4/(1/2)=8$. The same calculation also gives $\lim_{t\to\infty}4g(t)^2/t=8$, which we will use for the bounded-range limit below.

\textit{Ratio monotonicity and sharp bounds.}
For $0<x<y$, strict convexity and $F_{\mathrm{zCDP}}(0)=0$ give $F_{\mathrm{zCDP}}(x)<\frac{x}{y}F_{\mathrm{zCDP}}(y)$. Thus, $F_{\mathrm{zCDP}}(\rho)/\rho$ is strictly increasing, and hence so is its positive square root $\mu_{\mathrm{zCDP}}(\rho)/\sqrt{\rho}$. Its endpoint limits then give $\sqrt{\pi\rho}<\mu_{\mathrm{zCDP}}(\rho)<\sqrt{8\rho}$ for $\rho>0$, and both sides equal zero at $\rho=0$. Both constants are sharp by those limits. This completes all zCDP assertions.
\end{proof}

\brconversionconvexity*

\begin{figure}[t]
    \centering
    \includegraphics[scale=0.89]{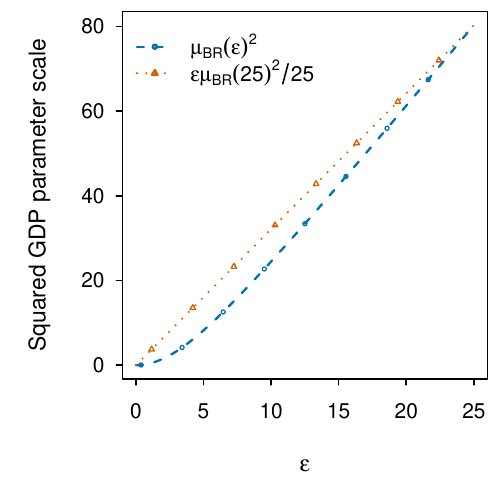}
    \caption{The squared sharp conversion $\varepsilon\mapsto\mu_{\mathrm{BR}}(\varepsilon)^2$ from logistic DP to GDP is strictly convex. The dotted line $\varepsilon\,\mu_{\mathrm{BR}}(25)^2/25$ is the chord from $0$ to $25$, and the curve lies below it.}
    \label{fig:muBR2}
\end{figure}

\begin{proof}[Proof of the Theorem~\ref{thm:br-conversion-shape}]
\textit{Strict convexity of the squared conversion.} 
Write $F_{\mathrm{BR}}(\varepsilon)=\mu_{\mathrm{BR}}(\varepsilon)^2$. This function is continuous and strictly increasing from zero to infinity, with $F_{\mathrm{BR}}(0)=0$. Solving $\sqrt{s}=2\Phi^{-1}(\sigma(\varepsilon/2))$ gives $F_{\mathrm{BR}}^{-1}(s)=2\log\frac{\Phi(\sqrt{s}/2)}{\Phi(-\sqrt{s}/2)}=2t(s)$. Using $J'(s)=\tanh(t(s)/2)t'(s)$ from the proof of Theorem~\ref{thm:conversion-shape}, we obtain
\begin{equation*}
(F_{\mathrm{BR}}^{-1})'(s)=\frac{2J'(s)}{\tanh(t(s)/2)}.
\end{equation*}
The numerator is positive and strictly decreasing; the denominator is positive and strictly increasing. The inverse slope is therefore positive and strictly decreasing. The inverse is strictly increasing and strictly concave, including its continuous extension to zero by the endpoint argument used above. Hence $F_{\mathrm{BR}}$ is strictly increasing and strictly convex on $[0,\infty)$.

\textit{Concavity of the unsquared conversion.}
Lemma~\ref{lem:quantile-ratio} gives $(g'/w)'\le0$. For $t>0$, $w'(t)<0$ and $g'/w>0$, so differentiating $g'=(g'/w)w$ yields $g''(t)<0$. Since $\mu_{\mathrm{BR}}(\varepsilon)=2g(\varepsilon/2)$, $\mu_{\mathrm{BR}}'(\varepsilon)=g'(\varepsilon/2)>0$ and $\mu_{\mathrm{BR}}''(\varepsilon)=\tfrac12g''(\varepsilon/2)<0$ for $\varepsilon>0$. Continuity extends strict concavity to $[0,\infty)$.

\textit{The two limits.}
The bounded-range conversion is $\mu_{\mathrm{BR}}(\varepsilon)=2g(\varepsilon/2)$. Thus the limits $\varepsilon\downarrow0$ and $\varepsilon\to\infty$ correspond directly to $t=\varepsilon/2\downarrow0$ and $t\to\infty$. We can use the Gaussian quantile calculations above without repeating their derivations.

\emph{As $\varepsilon\downarrow0$.}
Substitute $t=\varepsilon/2$ into $2g(t)=\sqrt{\pi/2}\,t+o(t)$ to obtain $\mu_{\mathrm{BR}}(\varepsilon)=\sqrt{\pi/8}\,\varepsilon+o(\varepsilon)$. Squaring gives $F_{\mathrm{BR}}(\varepsilon)=(\pi/8)\varepsilon^2+o(\varepsilon^2)$. Consequently, $\lim_{\varepsilon\downarrow0}\mu_{\mathrm{BR}}(\varepsilon)/\varepsilon
=\sqrt{\pi/8}$ and $\lim_{\varepsilon\downarrow0}F_{\mathrm{BR}}(\varepsilon)/\varepsilon=0$.

\emph{As $\varepsilon\to\infty$.}
The zCDP limit calculation established $\lim_{t\to\infty}4g(t)^2/t=8$. Since the bounded-range budget is $2t$, it follows that
\[
\lim_{\varepsilon\to\infty}\frac{F_{\mathrm{BR}}(\varepsilon)}{\varepsilon}
=\frac12\lim_{t\to\infty}\frac{4g(t)^2}{t}=4.
\]

\textit{Ratio monotonicity.}
For $0<x<y$, strict convexity gives $F_{\mathrm{BR}}(x)<(x/y)F_{\mathrm{BR}}(y)$, whereas strict concavity gives $\mu_{\mathrm{BR}}(x)>(x/y)\mu_{\mathrm{BR}}(y)$. Hence $\mu_{\mathrm{BR}}(\varepsilon)^2/\varepsilon$ strictly increases and $\mu_{\mathrm{BR}}(\varepsilon)/\varepsilon$ strictly decreases. The ratio $\mu_{\mathrm{BR}}(\varepsilon)^2/\varepsilon$ tends to $0$ as $\varepsilon\downarrow0$ and to $4$ as $\varepsilon\to\infty$. These monotonicities and the limits $\sqrt{\pi/8}$ and $4$ give $\mu_{\mathrm{BR}}(\varepsilon)/\varepsilon<\sqrt{\pi/8}$ and $\mu_{\mathrm{BR}}(\varepsilon)^2/\varepsilon<4$ for $\varepsilon>0$, and hence $\mu_{\mathrm{BR}}(\varepsilon)<\min\{\sqrt{\pi/8}\,\varepsilon,\,2\sqrt{\varepsilon}\}$. This completes the proof.
\end{proof}

\convertbeforecomposing*

\begin{figure}[t]
    \centering
    \includegraphics[scale=0.89]{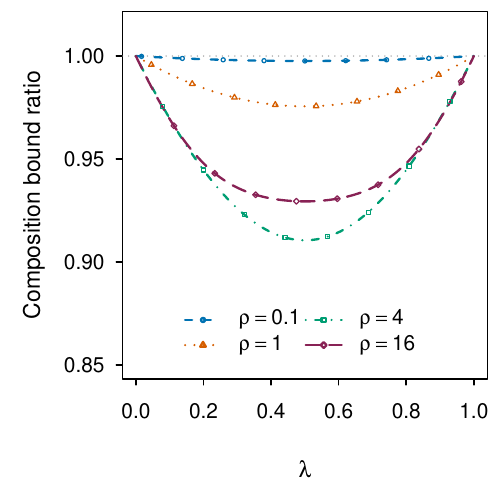}
    \includegraphics[scale=0.89]{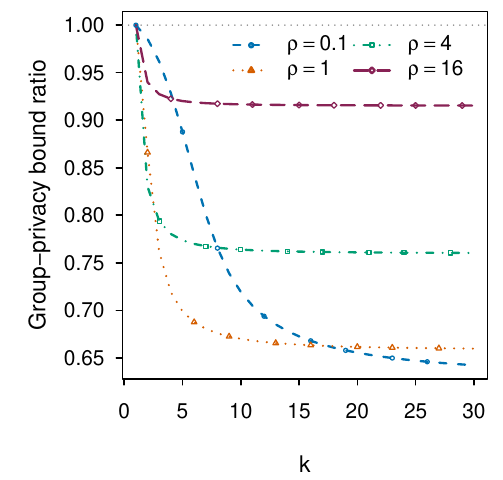}
    \caption{Left: composition shows $\sqrt{\mu_{\mathrm{zCDP}}(\lambda\rho)^2+\mu_{\mathrm{zCDP}}((1-\lambda)\rho)^2}
/\mu_{\mathrm{zCDP}}\!\left(\rho\right)\le1$. Right: group privacy shows $k\mu_{\mathrm{zCDP}}(\rho)/\mu_{\mathrm{zCDP}}(k^2\rho)\le1$.}
    \label{fig:zcdp-accounting}
\end{figure}

\brconvertbeforecomposing*

\begin{figure}[t]
    \centering
    \includegraphics[scale=0.89]{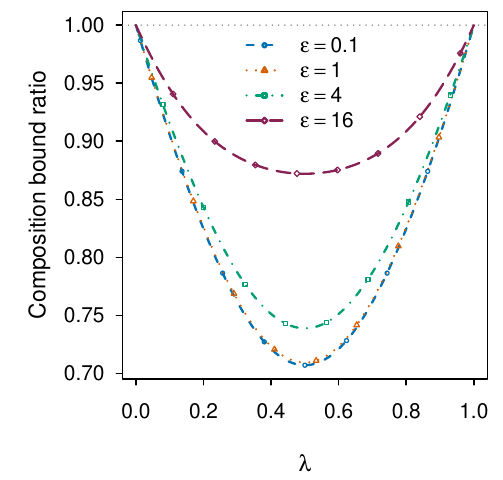}
    \includegraphics[scale=0.89]{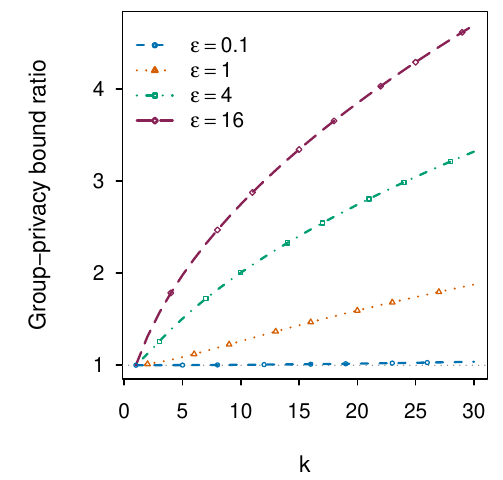}
    \caption{Left: composition shows $\sqrt{\mu_{\mathrm{BR}}(\lambda\varepsilon)^2+\mu_{\mathrm{BR}}((1-\lambda)\varepsilon)^2}
/\mu_{\mathrm{BR}}\!\left(\varepsilon\right)\le1$. Right: group privacy shows $k\mu_{\mathrm{BR}}(\varepsilon)/\mu_{\mathrm{BR}}(k\varepsilon)\ge1$.}
    \label{fig:br-accounting}
\end{figure}

\begin{proof}[Proof of the first inequalities in Corollaries~\ref{cor:composition-comparison} and~\ref{cor:br-composition-comparison}]
Let $F$ be either $F_{\mathrm{zCDP}}=\mu_{\mathrm{zCDP}}^2$ or $F_{\mathrm{BR}}=\mu_{\mathrm{BR}}^2$, both of which are strictly convex on $[0,\infty)$ with $F(0)=0$. For nonnegative $s_1,\ldots,s_n$, let $s=\sum_i s_i$. If $s=0$, both sides equal zero. Otherwise, writing $s_i=(s_i/s)s+(1-s_i/s)\cdot0$, convexity gives $F(s_i)\le(s_i/s)F(s)$. Summing over $i$ gives $\sum_iF(s_i)\le F(s)$, and taking square roots gives the first inequality of each corollary, with $s_i=\rho_i$ or $s_i=\varepsilon_i$. If at least two $s_i$ are positive, each positive $s_i$ lies strictly between $0$ and $s$, so strict convexity gives $F(s_i)<(s_i/s)F(s)$ for these $i$, and the inequality is strict.
\end{proof}

\begin{proof}[Proof of the second inequalities in Corollaries~\ref{cor:composition-comparison} and~\ref{cor:br-composition-comparison}]
If $k=1$, or if $\rho=0$ or $\varepsilon=0$, both sides are equal. Let $k>1$. For zCDP with $\rho>0$, write $\rho=k^{-2}(k^2\rho)+(1-k^{-2})\cdot0$. Strict convexity of $F_{\mathrm{zCDP}}$ and $F_{\mathrm{zCDP}}(0)=0$ give $F_{\mathrm{zCDP}}(\rho)<k^{-2}F_{\mathrm{zCDP}}(k^2\rho)$, and taking square roots gives $k\mu_{\mathrm{zCDP}}(\rho)<\mu_{\mathrm{zCDP}}(k^2\rho)$. For logistic DP with $\varepsilon>0$, write $\varepsilon=k^{-1}(k\varepsilon)+(1-k^{-1})\cdot0$. Strict concavity of $\mu_{\mathrm{BR}}$ and $\mu_{\mathrm{BR}}(0)=0$ give $\mu_{\mathrm{BR}}(\varepsilon)>k^{-1}\mu_{\mathrm{BR}}(k\varepsilon)$, that is, $\mu_{\mathrm{BR}}(k\varepsilon)<k\mu_{\mathrm{BR}}(\varepsilon)$.
\end{proof}

\section{Minimax comparison between zCDP and GDP}\label{app:mechanism-classes-minimax}
Let $\mathcal P$ be a class of distributions on $\mathcal X$, let $\theta:\mathcal P\to\Theta$ be the parameter of interest, let $d$ be a metric on $\Theta$, and let $\ell:[0,\infty)\to[0,\infty)$ be nondecreasing with $\ell(0)=0$. Given $n$ i.i.d.\ observations $X_{1:n}$ from $P\in\mathcal P$, a mechanism $Q$ produces $Z\sim Q(\cdot\mid X_{1:n})$, with the neighboring relation of Section~\ref{sec:privacy-preliminaries}. For a fixed mechanism $Q$, define $\mathfrak M_n(\theta(\mathcal P),\ell\circ d;Q):=\inf_{\widehat\theta}\sup_{P\in\mathcal P}\mathbb E_{P,Q}[\ell(d(\widehat\theta(Z),\theta(P)))]$, where the infimum is over estimators based only on the privatized output $Z$. Define the private minimax risks by
\begin{align*}
\mathfrak M_n^{\mathrm{GDP}}(\theta(\mathcal P),\ell\circ d;\mu)&:=\inf\{\mathfrak M_n(\theta(\mathcal P),\ell\circ d;Q):Q\text{ is }\mu\text{-GDP}\},\\
\mathfrak M_n^{\mathrm{zCDP}}(\theta(\mathcal P),\ell\circ d;\rho)&:=\inf\{\mathfrak M_n(\theta(\mathcal P),\ell\circ d;Q):Q\text{ is }\rho\text{-zCDP}\}.
\end{align*}
Since both guarantees are preserved under post-processing, they equal the minimax risks over estimators that themselves satisfy $\mu$-GDP or $\rho$-zCDP.

\begin{corollary}[Minimax comparisons with GDP]\label{cor:minimax-comparisons}
For every statistical problem $(\mathcal P,\theta,d,\ell)$, every $n\ge 1$, and every $\rho>0$,
\begin{align*}
\mathfrak M_n^{\mathrm{GDP}}(\theta(\mathcal P),\ell\circ d;\sqrt{8\rho})&\le\mathfrak M_n^{\mathrm{GDP}}(\theta(\mathcal P),\ell\circ d;\mu_{\mathrm{zCDP}}(\rho))\\
&\le\mathfrak M_n^{\mathrm{zCDP}}(\theta(\mathcal P),\ell\circ d;\rho)\le\mathfrak M_n^{\mathrm{GDP}}(\theta(\mathcal P),\ell\circ d;\sqrt{2\rho}).
\end{align*}
\end{corollary}

\begin{proof}
The infimum of the same worst-case risk can only decrease when it is taken over more mechanisms. Every $\sqrt{2\rho}$-GDP mechanism is $\rho$-zCDP by Proposition~\ref{prop:gdp-zcdp}, which gives the last inequality. Every $\rho$-zCDP mechanism is $\mu_{\mathrm{zCDP}}(\rho)$-GDP by Theorem~\ref{thm:sharp-gdp-conversions}, which gives the middle inequality. Finally, $\mu_{\mathrm{zCDP}}(\rho)<\sqrt{8\rho}$ by Theorem~\ref{thm:conversion-shape}, and $\mu$-GDP implies $\nu$-GDP for every $\nu\ge\mu$ because $G_\mu$ is decreasing in $\mu$, which gives the first inequality.
\end{proof}

The statistical problem and sample size are fixed throughout. Thus any risk upper bound achieved under $\sqrt{2\rho}$-GDP also holds under $\rho$-zCDP, and any minimax lower bound under $\mu_{\mathrm{zCDP}}(\rho)$-GDP, or under $\sqrt{8\rho}$-GDP, applies under $\rho$-zCDP. The comparison does not depend on the method used to prove a lower bound.

\section{Background on logistic differential privacy}
\label{app:logistic-background}

Let $Z$ be the standard logistic distribution, with CDF
$\sigma(z)=(1+e^{-z})^{-1}$ and density
\[
r(z)
=\frac{e^{-z}}{(1+e^{-z})^2}
=\frac{1}{4\cosh^2(z/2)}.
\]
The density is symmetric and strictly log-concave because $\frac{d^2}{dz^2}\log r(z)=-\frac12\operatorname{sech}^2(z/2)<0$. Log-concavity implies that the location family has monotone likelihood ratios. Thus a most powerful test of $Z$ against $Z+\varepsilon$ rejects
for large observations. At type~I error $\alpha\in(0,1)$, its
threshold is $c=\sigma^{-1}(1-\alpha)$, giving
\[
T(Z,Z+\varepsilon)(\alpha)
=\Pr(Z+\varepsilon\le c)
=\sigma\!\left(\sigma^{-1}(1-\alpha)-\varepsilon\right)
=L_\varepsilon(\alpha).
\]
This is the location-shift representation used in the proof of \cite[Proposition~7.2.2]{dong2020gaussian}. More generally, the relationship between log-concave noise distributions and functional composition is developed in \cite[Section~3]{awan2022logconcave}.

For trade-off functions $f=T(P,Q)$ and $g=T(P',Q')$, we define their tensor product as $f\otimes g:=T(P\times P',Q\times Q')$. By the adaptive composition theorem, if each release has conditional guarantee $f_i$ for every fixed realization of preceding outputs, the joint release satisfies $(f_1\otimes\cdots\otimes f_m)$-DP \cite[Theorem~4]{dong2022gaussian}. In contrast, group privacy for an $f$-DP mechanism is governed by functional composition $f^{[k]}:=1-(1-f)^{\circ k}$, where $\circ k$ denotes $k$-fold function composition \cite[Theorem~3]{dong2022gaussian}. For the logistic family, define the corresponding power bound
\[
h_\varepsilon(\alpha):=1-L_\varepsilon(\alpha)
=\frac{e^\varepsilon\alpha}
       {1-\alpha+e^\varepsilon\alpha}
=\sigma\!\left(\sigma^{-1}(\alpha)+\varepsilon\right).
\]
Its location-shift representation gives $h_b\circ h_a=h_{a+b}$, for $a,b\ge0$. Consequently, the general group-privacy formula specializes to $1-(1-L_\varepsilon)^{\circ k} =1-h_\varepsilon^{\circ k} =L_{k\varepsilon}$. An $\varepsilon$-logistic DP mechanism therefore satisfies
$k\varepsilon$-logistic DP for groups of size $k$.

For composition, the comparison between tensor and functional composition for symmetric trade-off functions \cite[Lemma~A.2]{awan2022logconcave}, translated into our decreasing trade-off convention, yields
\[
L_a\otimes L_b
\ge 1-(1-L_b)\circ(1-L_a)
=1-h_b\circ h_a
=L_{a+b}.
\]
Iterating this inequality gives $L_{\varepsilon_1}\otimes\cdots\otimes L_{\varepsilon_m}\ge L_{\sum_{i=1}^m\varepsilon_i}$. Consequently, if each release in an adaptive composition is $\varepsilon_i$-logistic DP for every fixed realization of the preceding outputs, with deterministic $\varepsilon_i$, then the joint release is $(\sum_{i=1}^m\varepsilon_i)$-logistic DP. 

Finally, the class of $\varepsilon$-BR mechanisms is not closed under mixtures, whereas a fixed $f$-DP class is closed under mixtures with dataset-independent weights \cite[Section~7.3]{dong2020gaussian}. Consequently, mixtures of $\varepsilon$-logistic DP mechanisms remain $\varepsilon$-logistic DP.

\clearpage

\bibliographystyle{alpha}
\bibliography{bibliography}
\end{document}